\documentclass[twocolumn]{aastex63}

\usepackage{amsmath}

\newcommand{\msun}{\hbox{M$_{\odot}$}}

\newcommand{\halpha}{\hbox{H$\alpha$}}

\newcommand{\swift}{\textit{Swift }}
\newcommand{\swiftn}{\textit{Swift}}

\graphicspath{{./}{figures/}}

\begin{document}

\title{On the Double: Two Luminous Flares from the Nearby Tidal Disruption Event ASASSN-22ci (AT2022dbl) and Connections to Repeating TDE Candidates}
\shorttitle{Two Flares of the TDE ASASSN-22ci (AT2022dbl)}


\shortauthors{Hinkle et al.}

\correspondingauthor{Jason T. Hinkle}
\email{jhinkle6@hawaii.edu}

\author[0000-0001-9668-2920]{Jason T. Hinkle}
\altaffiliation{NASA FINESST FI}
\affiliation{Institute for Astronomy, University of Hawai\`{}i at Manoa, 2680 Woodlawn Dr., Honolulu, HI 96822}
\affiliation{Kavli Institute for Theoretical Physics, University of California, Santa Barbara, CA 93106, USA}

\author[0000-0002-4449-9152]{Katie Auchettl}
\affiliation{School of Physics, The University of Melbourne, VIC 3010, Australia}
\affiliation{Department of Astronomy and Astrophysics, University of California, Santa Cruz, CA 95064, USA}

\author[0000-0003-3953-9532]{Willem B. Hoogendam}
\altaffiliation{NSF Fellow}
\affiliation{Institute for Astronomy, University of Hawai\`{}i at Manoa, 2680 Woodlawn Dr., Honolulu, HI 96822}

\author[0000-0003-3490-3243]{Anna V. Payne}
\affiliation{Space Telescope Science Institute, 3700 San Martin Drive, Baltimore, MD 21218, USA}

\author[0000-0001-9206-3460]{Thomas~W.-S.~Holoien}
\affiliation{Institute for Astronomy, University of Hawai\`{}i at Manoa, 2680 Woodlawn Dr., Honolulu, HI 96822}

\author[0000-0003-4631-1149]{Benjamin J. Shappee}
\affiliation{Institute for Astronomy, University of Hawai\`{}i at Manoa, 2680 Woodlawn Dr., Honolulu, HI 96822}
\affiliation{Kavli Institute for Theoretical Physics, University of California, Santa Barbara, CA 93106, USA}

\author[0000-0002-2471-8442]{Michael A. Tucker}
\altaffiliation{CCAPP Fellow}
\affiliation{Department of Astronomy, The Ohio State University, 140 West 18th Avenue, Columbus, OH 43210, USA}
\affiliation{Center for Cosmology and AstroParticle Physics, The Ohio State University, 191 W.\ Woodruff Ave., Columbus, OH 43210, USA}

\author[0000-0001-6017-2961]{Christopher~S.~Kochanek}
\affiliation{Department of Astronomy, The Ohio State University, 140 West 18th Avenue, Columbus, OH 43210, USA}
\affiliation{Center for Cosmology and AstroParticle Physics, The Ohio State University, 191 W.\ Woodruff Ave., Columbus, OH 43210, USA}

\author{K.~Z.~Stanek}
\affiliation{Department of Astronomy, The Ohio State University, 140 West 18th Avenue, Columbus, OH 43210, USA}
\affiliation{Center for Cosmology and AstroParticle Physics, The Ohio State University, 191 W.\ Woodruff Ave., Columbus, OH 43210, USA}

\author[0000-0001-5661-7155]{Patrick J. Vallely}
\affiliation{Core Pricing Systems, CarMax, Inc., Richmond, VA 23238, USA}

\author[0000-0002-4269-7999]{Charlotte R. Angus}
\affiliation{Astrophysics Research Centre, School of Mathematics and Physics, Queen's University Belfast, Belfast BT7 1NN, UK}

\author[0000-0002-5221-7557]{Chris Ashall}
\affiliation{Department of Physics, Virginia Tech, 850 West Campus Drive, Blacksburg VA, 24061, USA}
\affiliation{Institute for Astronomy, University of Hawai\`{}i at Manoa, 2680 Woodlawn Dr., Honolulu, HI 96822}

\author[0000-0001-6069-1139]{Thomas de Jaeger}
\affiliation{CNRS/IN2P3 (Sorbonne Université, Université Paris Cité), Laboratoire de Physique Nucléaire et de Hautes Énergies, 75005 Paris, France}

\author[0000-0002-2164-859X]{Dhvanil D. Desai}
\affiliation{Institute for Astronomy, University of Hawai\`{}i at Manoa, 2680 Woodlawn Dr., Honolulu, HI 96822}

\author[0000-0003-3429-7845]{Aaron Do}
\affiliation{Institute of Astronomy and Kavli Institute for Cosmology, Madingley Road, Cambridge, CB3 0HA, UK}

\author[0000-0002-9113-7162]{Michael M. Fausnaugh}
\affiliation{Department of Physics \& Astronomy, Texas Tech University, Lubbock, TX 79410-1051, USA}

\author[0000-0003-1059-9603]{Mark E. Huber}
\affiliation{Institute for Astronomy, University of Hawai\`{}i at Manoa, 2680 Woodlawn Dr., Honolulu, HI 96822}

\author[0000-0001-9719-4080]{Ryan J. Rickards Vaught}
\affiliation{Space Telescope Science Institute, 3700 San Martin Drive, Baltimore, MD 21218, USA}

\author[0009-0008-3724-1824]{Jennifer Shi}
\affiliation{School of Physics, The University of Melbourne, VIC 3010, Australia}

\begin{abstract}
\noindent We present observations of ASASSN-22ci (AT2022dbl), a nearby tidal disruption event (TDE) discovered by the All-Sky Automated Survey for Supernovae (ASAS-SN) at a distance of d$_L \simeq 125$ Mpc. Roughly two years after the initial ASAS-SN discovery, a second flare was detected coincident with ASASSN-22ci. UV/optical photometry and optical spectroscopy indicate that both flares are likely powered by TDEs. The striking similarity in flare properties suggests that these flares result from subsequent disruptions of the same star. Each flare rises on a timescale of $\sim$30 days, has a temperature of $\approx$30,000 K, a peak bolometric luminosity of $L_{UV/Opt} = 10^{43.6 - 43.9} \textrm{ erg} \textrm{ s}^{-1}$, and exhibits a blue optical spectrum with broad H, He, and N lines. No X-ray emission is detected during either flare, but X-ray emission with an unabsorbed luminosity of $L_{X} = 3\times10^{41} \textrm{ erg} \textrm{ s}^{-1}$ and $kT = 0.042$ eV is observed between the flares. Pre-discovery survey observations rule out the existence of earlier flares within the past $\approx$6000 days, indicating that the discovery of ASASSN-22ci likely coincides with the first flare. If the observed flare separation of $720 \pm 4.7$ days is the orbital period, the next flare of ASASSN-22ci should occur near MJD 61075 (2026 February 04). Finally, we find that the existing sample of repeating TDE candidates is consistent with Hills capture of a star initially in a binary with a total mass between $\sim$$1\mbox{ -- }4$ M$_{\odot}$ and a separation of $\sim$$0.01\mbox{ -- }0.1$ AU. 
\end{abstract}


\section{Introduction}

A tidal disruption event (TDE) occurs when a star passes so close to a supermassive black hole (SMBH) that the self-gravity of the star is overwhelmed by the tidal forces of the SMBH \citep[e.g.,][]{rees88, evans89, phinney89}. Many disruptions are likely to be ``full'', meaning that the star is totally disrupted with no bound stellar remnant. Nevertheless, as there is a larger cross-section for partial disruptions than full disruptions, partial TDEs are thought to be more common than full TDEs \citep[e.g.,][]{stone20, zhong22}. Partial TDEs are predicted to have faster declines than full TDEs \citep{coughlin19, miles20}, a lower peak fallback rate that scales with the impact parameter of the tidal encounter \citep[e.g.,][]{miles20, law-smith20}, and a delayed peak relative to the disruption \citep[e.g.,][]{miles20, law-smith20}.

In recent years a number of TDEs have had properties consistent with partial events, either from light curve modeling \citep{mockler19, gomez20, nicholl20} or the observation of multiple flares \citep{payne21, wevers23, somalwar23, lin24}. As TDEs, the events with multiple or repeating flares are of particular interest as, (1) they are guaranteed to be partial disruptions since a bound remnant exists and (2) if the flare recurrence time is reasonably well known, they can be used to study the early phases of TDE emission in unprecedented detail \citep[e.g.,][]{payne22, payne23}. The multiple/repeating TDEs claimed to date are the optically-selected ASASSN-14ko \citep{payne21, payne22, payne23}, ASASSN-18ul \citep[AT2018fyk;][]{wevers19, wevers23}, ZTF20acaazkt \citep[AT2020vdq;][]{somalwar23}, and ASASSN-22ci \citep[AT2022dbl;][the object studied in this work]{lin24}, the X-ray-selected events eRASSt J045650.3–203750 \citep{liu23, liu24_eRASSt} and RX J133157.6-324319.7 \citep{malyali23}. There are also several candidate events in AGN hosts including ZTF19aaejtoy \citep[AT2019aalc;][]{veres24}, ZTF18aanlzzf \citep[AT2021aeuk;][]{sun25}, and Gaia22ahd \citep[AT2022agi;][in IRAS F01004-2237]{sun24}. Finally, the TDE ASASSN-19dj exhibited a nuclear flare in Catalina Real-Time Transient Survey \citep[CRTS;][]{drake09} data roughly 14.5 years before the flare presented in \citet{hinkle21a}. Given the growing sample of multiple/repeating TDEs since the discovery of ASASSN-19dj, it now seems likely that this earlier flare was also a TDE. These events have a wide range of flare recurrence times, between 115 days \citep{payne23} and $\sim$26 years \citep{malyali23}, and a diversity of behaviors across the electromagnetic spectrum \citep{payne23, lin24}.

The repeating partial TDEs are likely the result of tidal capture through the Hills mechanism \citep{hills88, cufari22, lu23}. In this scenario, a tight binary passes close to an SMBH and is disrupted, resulting in one star becoming tightly bound to the SMBH and the other leaving as a hyper-velocity star. If the star that remains bound to the SMBH has a pericenter near its tidal radius it may continue to be partially disrupted at each pericenter passage, possibly powering multiple luminous flares. This scenario also provides insight into the expected distribution of the orbital periods of such repeating systems, as the initial binary must have an orbit that is hard relative to the velocity dispersion in the center of the galaxy \citep[e.g.,][]{hills88, cufari22, bandopadhyay24}. This sets the largest expected orbital period for the captured star. For SMBH masses consistent with observed TDEs and a binary consisting of two Solar-mass stars, we would expect orbital periods of $\lesssim 15$ years following Hills capture \citep[e.g.,][]{hills88, cufari22, bandopadhyay24}.

With the notable exception of ASASSN-14ko \citep{payne21, payne22, payne23}, most multiple/repeating TDE candidates have only exhibited two flares to date. Therefore, it remains to be seen if these are truly repeating events. If these events repeat, future flares will provide stronger constraints on the orbital period of the now-bound star. Recent simulations suggest that subsequent partial disruptions should become increasingly luminous until the star is finally fully disrupted \citep{liu24_simulation}. In such a scenario, transient surveys will be biased towards detecting the later and more luminous flares.

Despite many well-studied TDEs from optical surveys \citep[for a review, see][]{gezari21}, the mechanism powering the observed UV/optical emission remains a matter of debate. Broadly, there are two proposed models: reprocessed emission from an accretion disc \citep[e.g.][]{dai18, thomsen22} or shocks from collisions of the tidal debris streams \citep[e.g.][]{lu20, ryu20, steinberg24, krolik24}. This question remains the largest obstacle towards using TDEs as probes of the masses and spins of otherwise-quiescent SMBHs \citep[e.g.,][]{mockler19, gafton19, mummery24a, mummery24b}. Repeating partial UV/optical TDEs are an ideal test bed for such models as we can observe all phases of a TDE in detail. In particular, being able to predict future outbursts allows us to conduct dedicated follow-up of the earliest phases of emission and the rise to peak \citep[e.g.,][]{payne22, payne23}, phases which have only been caught, in detail, for two non-repeating TDEs, ASASSN-19bt \citep{holoien19c} and AT2023lli \citep[][]{huang24}.

In this manuscript, we present the discovery and follow-up observations of the TDE ASASSN-22ci (AT2022dbl). Our analysis complements the earlier study of \citet{lin24}, with the notable additions of a high-cadence and high-precision light curve from the Transiting Exoplanet Survey Satellite \citep[TESS;][]{ricker15}, a detailed comparison of ASASSN-22ci to models of repeating partial TDEs, and an exploration of the binary parameters for the sample of multiple/repeating TDEs under the assumption of Hills capture. Throughout the paper we assume a cosmology of $H_0$ = 69.6 km s$^{-1}$ Mpc$^{-1}$, $\Omega_{M} = 0.29$, and $\Omega_{\Lambda} = 0.71$ \citep{wright06, bennett14}. The paper is organized in the following manner. In Section \ref{sec:discovery} we present the discovery of the transient and analyze its host galaxy properties. We detail our follow-up observations of the TDE in Section \ref{sec:data} and analyze these data in Section \ref{sec:analysis}. Section \ref{sec:multiple_flares} provides a comparison between the two flares seen for ASASSN-22ci and their context within the population of optically-selected multiple/repeating TDEs. Finally, we discuss our conclusions in Section \ref{sec:disc_conc}.

\section{Discovery and Host-Galaxy Properties} \label{sec:discovery}

\subsection{Transient Discovery} \label{sec:22ci_flare1}

ASASSN-22ci $(\alpha,\delta)=$ (12:20:45.010, $+$49:33:04.68) was discovered by the All-Sky Automated Survey for Supernovae \citep[ASAS-SN;][]{shappee14, kochanek17, hart23} in $g$-band data from the ASAS-SN ``Brutus'' unit on Haleakal\={a}, Hawai`i on 2022 February 22.6 UTC \citep{stanek22ci}. The discovery was promptly announced on the Transient Name Server (TNS) and given the TNS identification AT2022dbl\footnote{\url{https://wis-tns.weizmann.ac.il/object/2022dbl}}. For the remainder of this manuscript, we will use the discovery survey name, ASASSN-22ci.

ASASSN-22ci occurred in the nucleus of the galaxy WISEA J122045.05+493304.7, at a redshift of z = 0.0284 \citep{abazajian04}. This redshift corresponds to a luminosity distance of 125.0 Mpc. The SDSS spectrum of the host galaxy shows strong Balmer absorption and no strong emission lines, consistent with the post-starburst host galaxies often seen for TDEs \citep[e.g.,][]{arcavi14, french16}. 

Shortly before the ASAS-SN discovery, the Zwicky Transient Facility \citep[ZTF;][]{bellm19, masci19} also detected a brightening of this source \citep{arcavi_tnsnote}. This detection of the brightening of ASASSN-22ci in ZTF photometry was linked with the previously reported AT2018mac\footnote{\url{https://wis-tns.weizmann.ac.il/object/2018mac}}. Upon further inspection, it was clear that the detection reported in 2018 was erroneous and there was no transient at that time. Nevertheless, based on the early rise of ASASSN-22ci in ZTF photometry, the StarDestroyers team obtained an early-time spectrum with the FLOYDS spectrograph on the Faulkes Telescope North \citep{arcavi22}. This spectrum showed a strong blue continuum and broad emission lines consistent with \ion{He}{2} $\lambda 4686$ and H$\alpha$ at the host redshift. These spectroscopic features and a nuclear location within a likely post-starburst galaxy indicated that ASASSN-22ci was a TDE. Because of this, we triggered additional spectroscopic and photometric follow-up of ASASSN-22ci.

Much like a normal TDE, ASASSN-22ci faded to the flux of the host by approximately one year post-peak. Nevertheless, roughly two years after the initial discovery of ASASSN-22ci, a second brightening was detected by ZTF \citep{yao24} and later seen in Asteroid Terrestrial Impact Last Alert System \citep[ATLAS;][]{tonry18, smith20} and ASAS-SN light curves. This second flare was broadly similar to the first flare in terms of its light curve and exhibited a spectrum extremely similar to that of the first flare. This suggested that ASASSN-22ci is a multiple/repeating partial TDE, joining the small, but growing, class of such events \citep[e.g.,][]{payne21, wevers23, somalwar23, liu23}.

\subsection{Host-Galaxy Properties} \label{sec:archival}

\begin{figure*}
\centering
 \includegraphics[height=0.47\textwidth]{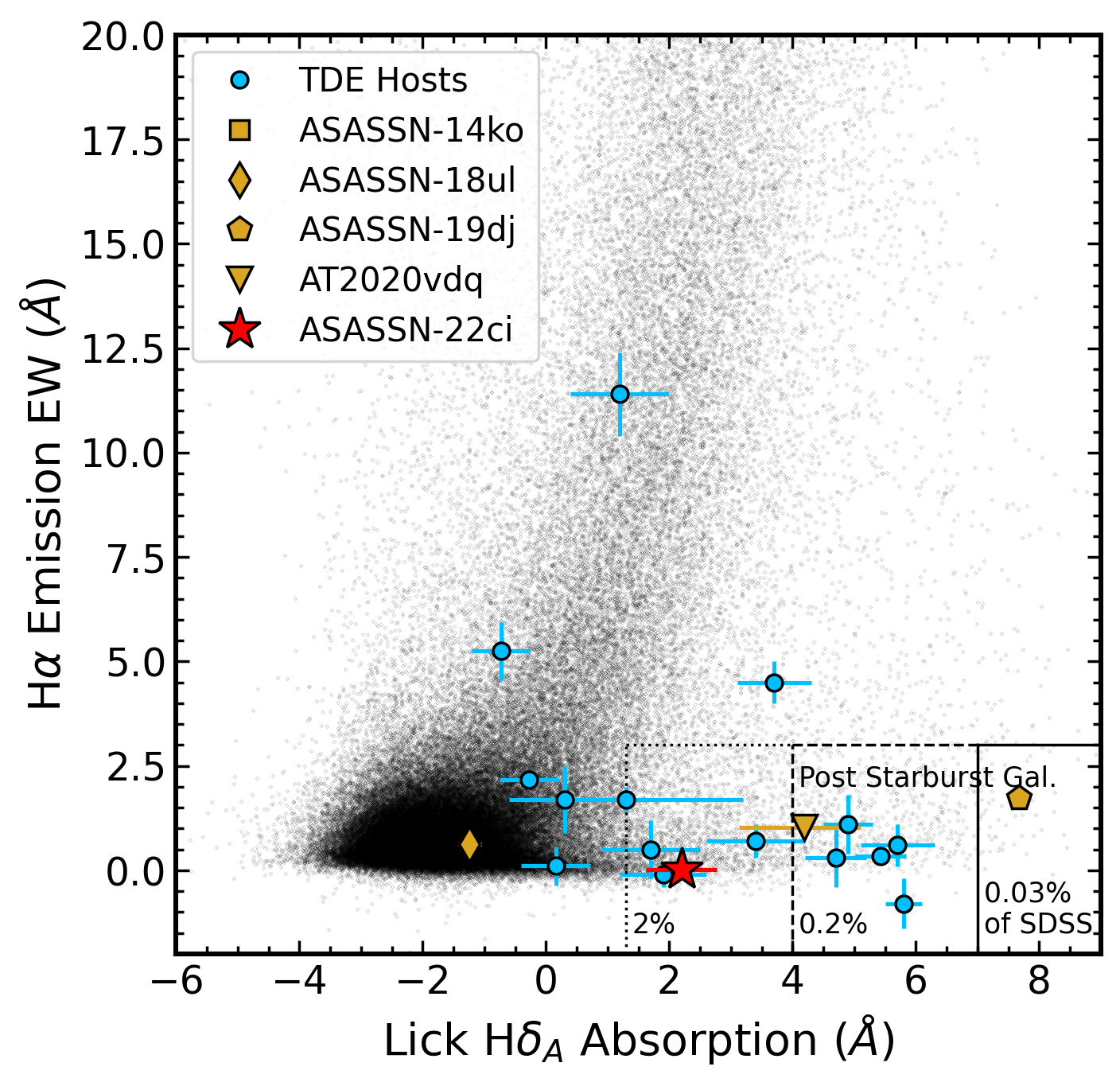}\hfill
 \includegraphics[height=0.47\textwidth]{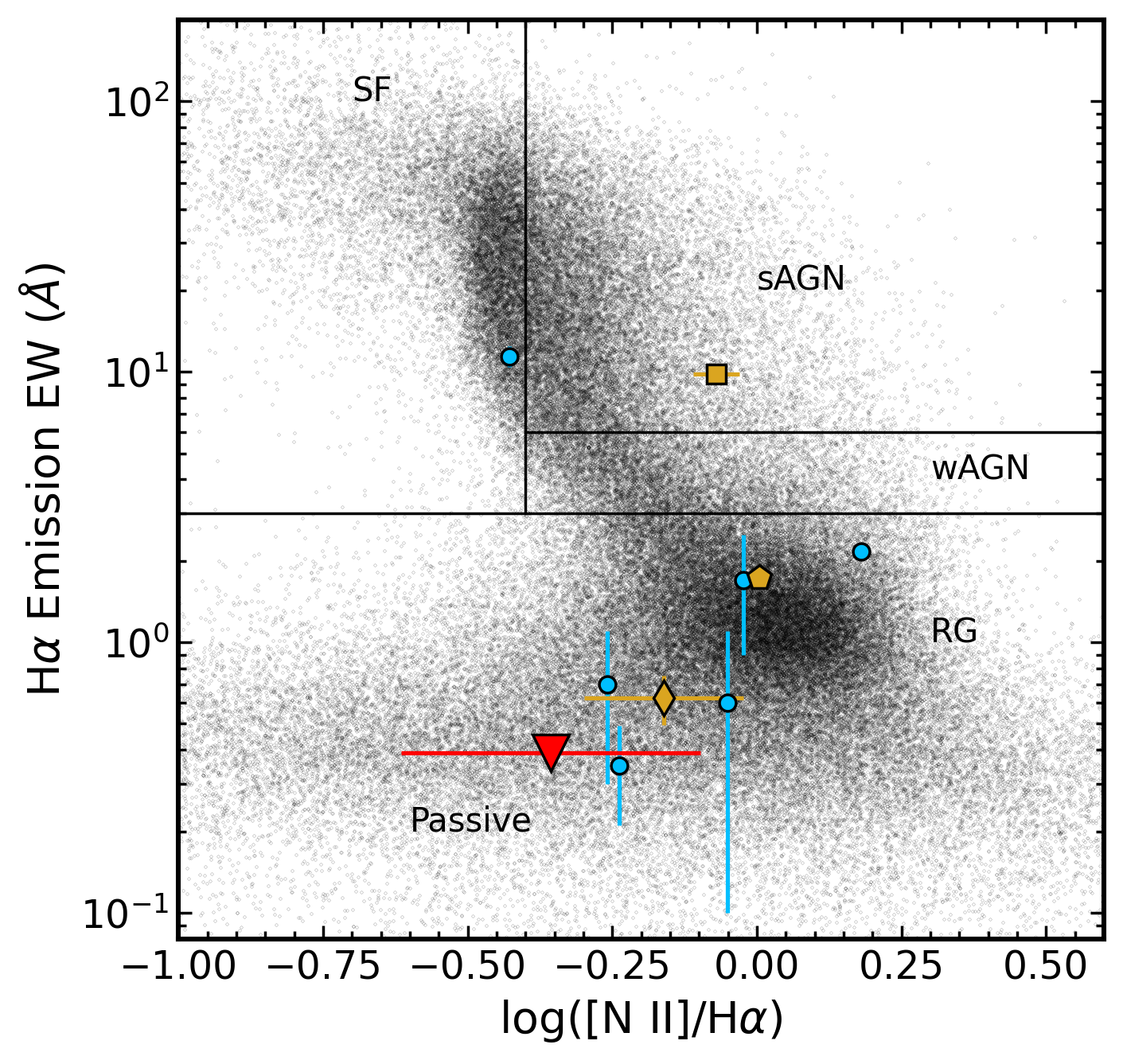} \\
 \caption{\textit{Left Panel}: The H$\alpha$ emission equivalent width (EW), tracing current star formation, as compared to the Lick H$\delta_A$ absorption index, tracing star formation within the past Gyr. The host galaxy of ASASSN-22ci is shown as a red symbol, a comparison sample of TDE hosts from \citet{graur18} is shown as blue circles, and the host galaxies of multiple/repeating TDEs are shown in gold. \textit{Right Panel}: The H$\alpha$ emission equivalent width (W$_{H\alpha}$) as compared to log$_{10}$([\ion{N}{2}] / \halpha), known as the WHAN diagram \citep{cidfernandes11}. Here, the downward-facing triangle for the host of ASASSN-22ci indicates a 3$\sigma$ upper limit. The lines separate star-forming galaxies (SF), strong AGN (sAGN), weak AGN (wAGN), passive, and ``retired galaxies'' (RG) \citep{cidfernandes11}. In both panels, we show galaxies from SDSS Data Release 8 \citep{eisenstein11} as gray background points.}
 \label{fig:ew_bpt}
\end{figure*}

The host galaxy WISEA J122045.05+493304.7 was well observed in the UV, optical, and IR prior to the discovery of ASASSN-22ci. We computed $NUV$ photometry of the host galaxy from Galaxy Evolution Explorer \citep[GALEX;][]{martin05} data using a 10\farcs{0} radius aperture and gPhoton \citep{million16}. We also obtained images in the $ugriz$ bands from the SDSS Data Release 15 \citep{aguado19}, $JHK_S$ images from the Two Micron All-Sky Survey \citep[2MASS;][]{skrutskie06}, and $W1,W2$ magnitudes from the Wide-field Infrared Survey Explorer \citep[WISE;][]{wright10} AllWISE survey. We measured 10\farcs{0} radius aperture magnitudes for the SDSS and 2MASS images, using several field stars to calibrate the magnitudes. For the AllWISE images, we also measured magnitudes in a 10\farcs{0} radius aperture, using the zero points listed in the image headers. The host-galaxy magnitudes are given in Table \ref{tab:arch_phot}.

\begin{deluxetable}{ccc}
\tablewidth{240pt}
\tabletypesize{\footnotesize}
\tablecaption{Archival Host Galaxy Photometry}
\tablehead{
\colhead{Filter} &
\colhead{AB Magnitude} &
\colhead{Magnitude Uncertainty}}
\startdata
$NUV$ & 21.03 & 0.08 \\
$u$ & 18.30 & 0.17 \\
$g$ & 16.59 & 0.08 \\
$r$ & 15.80 & 0.05 \\
$i$ & 15.47 & 0.04 \\
$z$ & 15.23 & 0.04 \\
$J$ & 15.04 & 0.06 \\
$H$ & 14.97 & 0.05 \\
$K_S$ & 14.99 & 0.06 \\
$W1$ & 15.94 & 0.02 \\
$W2$ & 16.62 & 0.02 \\
\enddata 
\tablecomments{Archival magnitudes of the host galaxy WISEA J122045.05+493304.7 used for our SED modeling. The NUV magnitude is a 10\farcs{0} aperture magnitude measured using gPhoton \citep{million16}. The $ugriz$, $JHK_S$, and $W1,W2$ magnitudes are measured from 10\farcs{0} aperture photometry performed on SDSS, 2MASS, and AllWISE images, respectively. All magnitudes are in the AB system using standard conversions.} 
\label{tab:arch_phot} 
\end{deluxetable}

We characterized the host galaxy using the Fitting and Assessment of Synthetic Templates \citep[\textsc{Fast};][]{kriek09} to fit stellar population synthesis models to the archival host-galaxy photometry. We assume a \citet{cardelli89} extinction law with $\text{R}_{\text{V}} = 3.1$ and foreground Galactic extinction of $\text{A}_{\text{V}} = 0.052$ mag \citep{schlafly11}, a Salpeter IMF \citep{salpeter55}, an exponentially declining star-formation rate, and the \citet{bruzual03} stellar population models. We find a stellar mass of M$_* = 9.3^{+2.1}_{-1.4} \times 10^9$~\msun, an age of $2.8^{+0.7}_{-0.9}$ Gyr, and a 3$\sigma$ upper-limit on the SFR of $<2.7 \times 10^{-2}$ M$_{\odot}$ yr$^{-1}$. From the M$_{BH}$ - M$_{*}$ relation of \citet{reines15}, we estimate a central SMBH mass of $10^{6.4}$ M$_{\odot}$, consistent with the estimate of \citet{lin24}. These host-galaxy physical parameters are typical of optically-selected TDEs \citep[e.g.,][]{hinkle21b, vanvelzen21, hammerstein23}.

We also used the archival SDSS \citep{york00} spectrum of WISEA J122045.05+493304.7 to place it in context with other TDE hosts. The spectrum shows no discernible emission lines and deep absorption in the Balmer series of hydrogen as well as other indicators of late-type stellar population such as Ca H and K and Mg I absorption. The spectrum is visually similar to those of other TDE hosts. We additionally obtain the measured line fluxes from the MPA-JHU catalog \citep{brinchmann04}. The results are shown in Figure \ref{fig:ew_bpt} compared to the broader sample of SDSS galaxies in black, a comparison sample of TDE hosts \citep[e.g.,][]{graur18} in blue, and the host galaxies of the multiple/repeating TDEs ASASSN-14ko \citep{payne21, tucker21}, ASASSN-18ul \citep{wevers19, wevers23}, ASASSN-19dj \citep{hinkle21a}, and AT2020vdqu \citep{somalwar23} in gold. 

For the multiple/repeating TDE hosts, we adopted host-galaxy measurements from the literature when possible. We do not include ASASSN-14ko in the left panel of Fig.\ \ref{fig:ew_bpt} as the MUSE spectrum of the host galaxy does not cover H$\delta$. For ASASSN-18ul and AT2020vdq we measured the Lick H$\delta_A$ absorption index using PyPHOT\footnote{\url{https://mfouesneau.github.io/docs/pyphot/\#}}. As AT2020vdq has no published host spectrum, we also estimated a 3$\sigma$ limit on the H$\alpha$ emission line equivalent width following \citet{leonard01} and assuming a conservative narrow line width of 300 km s$^{-1}$. We cannot estimate a log$_{10}$([\ion{N}{2}]$/$\halpha) ratio for AT2020vdq and therefore it is not shown in the right panel of Fig.\ \ref{fig:ew_bpt}.

The left panel of Fig.\ \ref{fig:ew_bpt} compares the H$\alpha$ emission line equivalent width (tracing current star formation) to the Lick H$\delta_A$ absorption index (tracing star formation within the past Gyr). The host of ASASSN-22ci has an H$\alpha$ emission EW of $0.02 \pm 0.13$ \AA{} and a Lick H$\delta_A$ index of $2.2 \pm 0.6$ \AA. As can be seen in Fig.\ \ref{fig:ew_bpt}, this identifies the host of ASASSN-22ci as a ``quiescent Balmer strong'' galaxy, living in a region where TDEs are over-represented relative to the underlying galaxy distribution \citep{french16, law-smith17}.

The right panel of Fig.\ \ref{fig:ew_bpt}, often called the WHAN diagram \citep{cidfernandes11}, compares the H$\alpha$ emission EW to the log$_{10}$([\ion{N}{2}]$/$\halpha). This parameter space separates galaxies by their dominant ionization mechanism. To be classified as an AGN, a galaxy must exhibit H$\alpha$ emission with an EW above 3 \AA. The log$_{10}$([\ion{N}{2}]$/$\halpha) ratio further differentiates between star formation and AGN ionization. Below the H$\alpha$ emission line EW limit, AGN ionization is unlikely to be dominant and the WHAN diagram uses the log$_{10}$([\ion{N}{2}]$/$\halpha) ratio to distinguish truly passive galaxies from the so-called ``retired galaxies.'' In this space, using a 3$\sigma$ upper-limit on the H$\alpha$ emission EW, the host of ASASSN-22ci is clearly passive, with no evidence for the presence of significant ionizing flux from either stars or an AGN.

We also use the WISE $W1 - W2$ color of the host to place limits on the existence of strong AGN activity \citep[e.g.,][]{assef10, assef13, stern12}. The  $W1 - W2$ color for the host of ASASSN-22ci is $-0.04 \pm 0.03$ Vega mag, fully consistent with a typical galaxy. Any AGN activity is significantly fainter than the stellar light in the MIR bands.

Pre-flare X-ray observations in the ROSAT All-Sky Survey (RASS) in 1990 and a ROSAT pointed Position Sensitive Proportional Counter (PSPC) observation of the nearby white dwarf PG 1218+497 in 1992, we find no significant emission associated with the host galaxy, with a 3$\sigma$ upper-limit of $<6\times10^{-3}$ counts s$^{-1}$ and $<1.5\times10^{-3}$ counts s$^{-1}$, respectively. Assuming a photon index of $\Gamma$ = 1.75 \citep[e.g.,][]{ricci17} and a Galactic column density of $N_{H}=1.94\times10^{20}$ cm$^{-2}$ along the line of sight \citep{HI4PI16}, this corresponds to an unabsorbed flux of $<2.0\times10^{13}$ erg cm$^{-2}$ s$^{-1}$ and $<5.1\times10^{13}$ erg cm$^{-2}$ s$^{-1}$ in the 0.3 - 10 keV band, respectively. These correspond to a X-ray luminosity limits of $<3.9\times10^{41}$ erg s$^{-1}$ and $<9.6\times10^{41}$ erg s$^{-1}$. With an upper limit on the Eddington ratio of $<0.1\%$, we again find that the host does not harbor a strong AGN \citep[e.g.,][]{ricci17}, although it remains consistent with a weak or low luminosity AGN.

\begin{figure*}
\centering
 \includegraphics[width=0.99\textwidth]{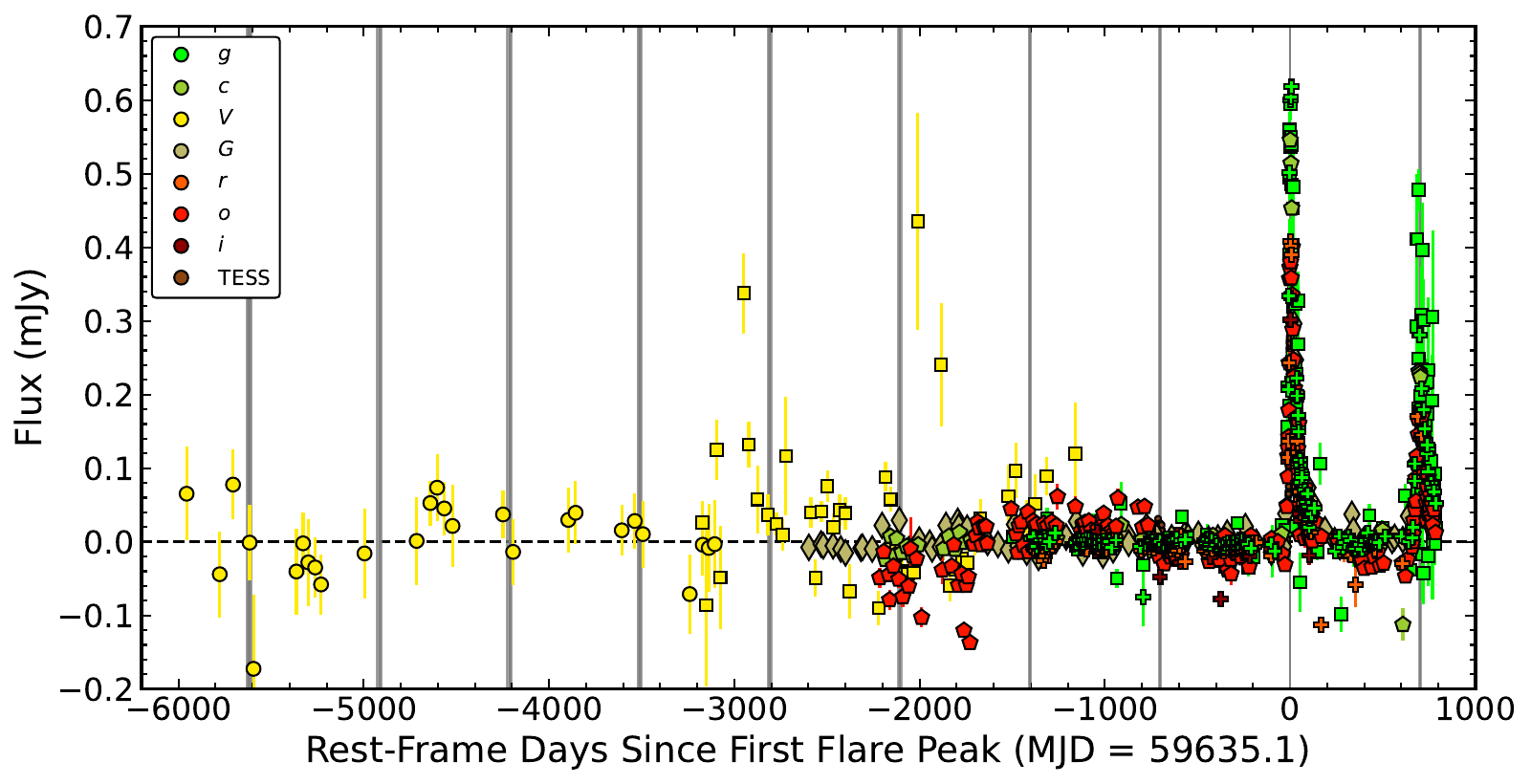}
 \caption{Long-term host-subtracted and foreground extinction-corrected light curve for the host galaxy of ASASSN-22ci. Shown are data from CRTS ($V$, yellow circles), ASAS-SN ($gV$, green and yellow squares), Gaia ($G$, khaki diamonds), ATLAS ($co$, yellow-green and red pentagons), ZTF ($gri$, green, orange, and dark red pluses), and TESS (brown octagons). The horizontal black dashed line represents zero flux. The vertical gray bands are the observed flare recurrence time projected into the past, with the width of the band representing the uncertainty on the time of the expected flare peak. We find no evidence of flares occurring at these epochs.}
 \label{fig:long_lc}
\end{figure*}

We also searched for previous nuclear variability using archival optical data. These included data from CRTS ($V$), ASAS-SN ($gV$), Gaia \citep[$G$;][]{hodgkin13}, ATLAS ($co$), ZTF ($gri$), and TESS. The available data are shown in Figure \ref{fig:long_lc}. We binned the CRTS data in monthly bins to increase the S/N per epoch and the Gaia data in 5-d bins. No significant variability is detected prior to the first flare of ASASSN-22ci. Furthermore, assuming that the $\sim$699 day rest-frame interval between the two observed flares is the flare recurrence time, we find no evidence for any prior activity at those epochs. There are two bright ASAS-SN $V$-band points roughly 2000 days before the first flare of ASASSN-22ci, close to one of the projected times of a possible earlier flare. These images are of moderate quality and have visible flat-fielding imperfections, making any detections less reliable. Additionally, the lack of any contemporaneous variability in the Gaia or ATLAS light curves indicates that no flare occurred at this time. Additionally, we examined the light curve around the time of the false positive detection of AT2018mac reported by ZTF. The discovery reported to TNS is MJD 58254.2 (a rest-frame phase of $\sim$$-1350$ day relative to the first flare peak). No flux excess is observed at this time, including in ZTF forced photometry.

\section{Observations of the TDE Flares} \label{sec:data}

\begin{figure*}
\centering
 \includegraphics[width=0.99\textwidth]{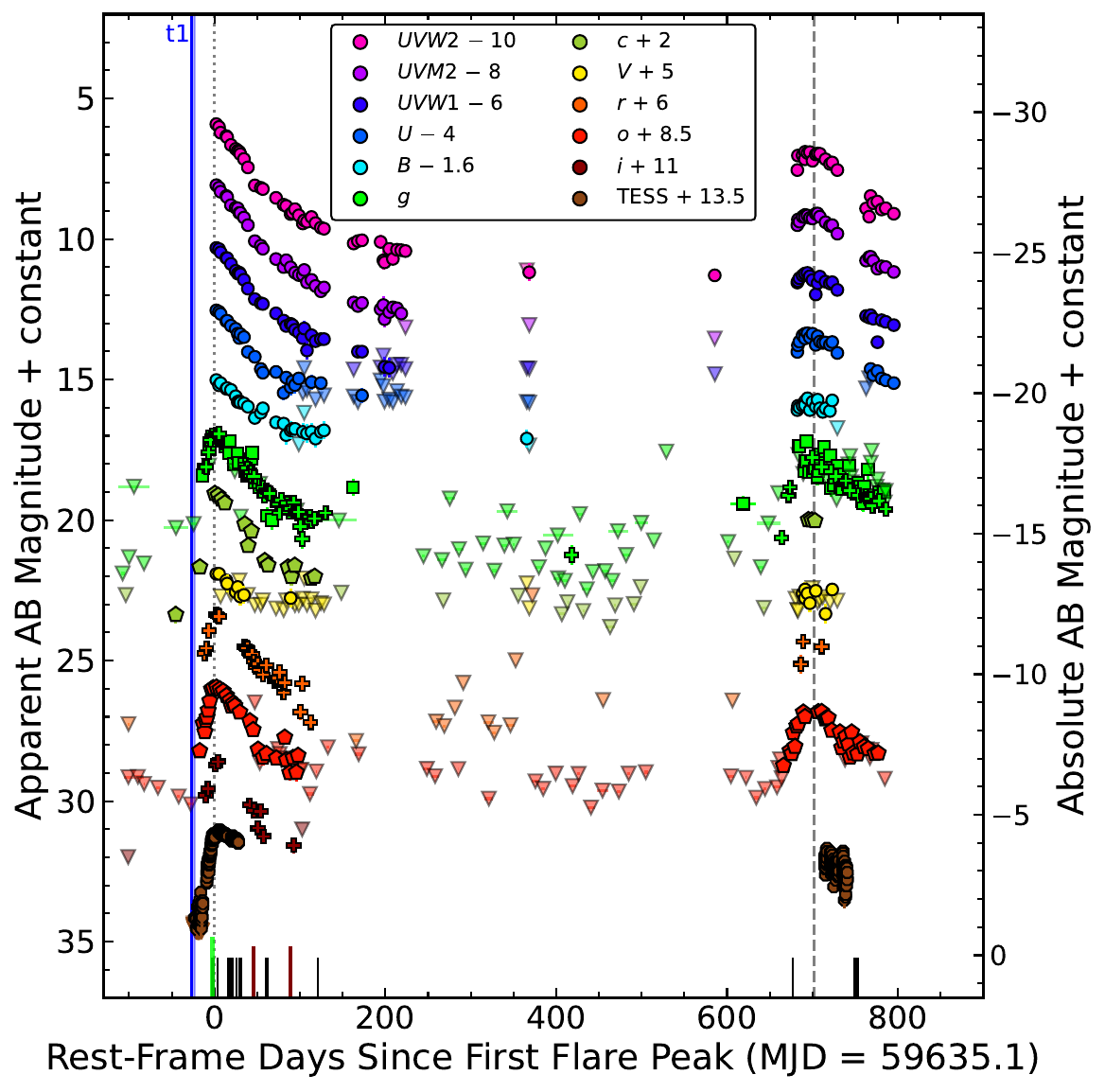}
 \caption{Host-subtracted and Galactic foreground extinction-corrected UV/optical light curves of ASASSN-22ci. We show data from Swift (UV+$UBV$; circles), ASAS-SN ($g$; squares), ATLAS ($co$; pentagons), ZTF ($gri$; pluses), and TESS (octagons). Generally, the survey photometry is stacked in 1-d bins during the two flares, and in 10-d bins (for ATLAS and ZTF) or 30-d bins (for ASAS-SN) outside of the flares. The TESS photometry is stacked in 2-h bins. Horizontal error bars indicate the range of observations over which the data has been stacked, although they are often smaller than the symbols. Downward-facing, translucent symbols are 3$\sigma$ upper limits for the filter of the same color. The vertical blue line represents the time of first light, with shading equal to the uncertainty. The green bar on the x-axis marks the epoch of ASAS-SN discovery, $\approx$2.5 rest-frame days before the peak of the first flare. The black (dark red) bars along the x-axis show epochs of optical (NIR) spectroscopic observations. The dotted (dashed) vertical line marks the peak time for the first (second) flare. All data are in the AB magnitude system.}
 \label{fig:opt_lc}
\end{figure*}

\subsection{ASAS-SN Light Curve}

ASAS-SN is an automated transient survey, currently comprising 20 telescopes on 5 robotic mounts. Each telescope is a 14-cm aperture Nikon telephoto lens with 8\farcs{0} pixels, with each unit hosting 4 telescopes. There are five ASAS-SN units, one each at the Haleakal\=a Observatory, the McDonald Observatory, and the South African Astrophysical Observatory (SAAO), and two at Cerro Tololo Inter-American Observatory (CTIO).

The images were reduced with the automated ASAS-SN pipeline, which utilizes the ISIS image subtraction package \citep{alard98, alard00}.  For the ASAS-SN $g$-band data, we rebuilt the reference using images taken outside of the two flares. The ASAS-SN $V$-band images show no source, so the reference image was created with typical procedures \citep[e.g.,][]{shappee14, kochanek17}. We then used the IRAF {\tt apphot} package with a 2-pixel radius ($\approx$16\farcs{0}) to perform aperture photometry, calibrating the photometry to the AAVSO Photometric All-Sky Survey \citep{Henden2015}. The pre-flare ASAS-SN $V$-band data, stacked in 30-d bins for higher S/N per epoch, is shown in Figure \ref{fig:long_lc} as yellow squares. We stacked the ASAS-SN $g$-band data during the flares in 1-d bins and data outside of the flares in 30-day bins for deeper limits. The ASAS-SN $g$-band photometry and 3$\sigma$ upper-limits are presented in Table \ref{tab:phot} and shown in Figures \ref{fig:long_lc} and \ref{fig:opt_lc} as green squares. 

\subsection{ATLAS Light Curve}

The ATLAS survey consists of four 0.5-m Wright Schmidt telescopes, one each on Haleakal\=a and Mauna Loa in Hawaii, one in El Sauce, Chile, and one in Sutherland, South Africa. ATLAS uses two broad-band filters, the `cyan' ($c$) filter from 4200--6500 \AA\ and the `orange' ($o$) filter from 5600--8200 \AA\ \citep{tonry18}. During typical operation, ATLAS takes a set of $\approx$4 images of a field over a short time period to track moving objects.

We obtained ATLAS $c$ and $o$ light curves from their forced point-spread function (PSF) photometry service \citep{shingles21}. For ATLAS data during the TDE flares, we stacked the intra-night epochs in 1-d bins to increase the S/N. Outside of the flares, we stacked the ATLAS data in 10-d bins to provide deeper limits on any source variability. We show the ATLAS $c$- and $o$-band photometry in Figures \ref{fig:long_lc} and \ref{fig:opt_lc} as yellow-green and red pentagons, respectively. We also present the ATLAS photometry and 3$\sigma$ upper-limits in Table \ref{tab:phot}. In the first two observing seasons, the ATLAS data appears to show large trends that are not present in other photometry, likely indicating they are systematic in nature. Nevertheless, these systematics are only present in data earlier than $\approx$$-900$ days relative to the first flare.

\subsection{ZTF Light Curve}

ZTF uses the Samuel Oschin 48" Schmidt telescope at Palomar Observatory. We obtained ZTF differential photometry in the $g$-, $r$-, and $i$-bands from their forced PSF photometry service \citep{masci19}. Similar to the ATLAS data, we stacked the intra-night photometric observations. The ZTF $g$-, $r$-, and $i$-band photometry is shown in Figures \ref{fig:long_lc} and \ref{fig:opt_lc} with green, orange, and dark-red pluses, respectively. The ZTF photometry and 3$\sigma$ upper limits are also given in Table \ref{tab:phot}.

\subsection{\textit{TESS} Observations}

ASASSN-22ci was observed by TESS during both flares (Sectors 48/49 and 75\footnote{Although the source position was observed by TESS in Sector 75, it is too close to the edge of the chip to yield a useful light curve.}/76), in addition to one pre-flare sector (Sector 22) from which we can place a strong constraint on AGN-like variability. To extract the TESS light curves we used the ISIS image subtraction package \citep{alard98, alard00} on the TESS full frame images (FFIs) following the procedures in \citet{vallely19, vallely21} and \citet{fausnaugh21}. For Sector 22, during the prime mission, we constructed reference images from the first 100 FFIs of good quality. For the extended mission sectors, we constructed reference images from the first 300 good-quality FFIs. We excluded images with sky background levels or PSF widths above the sector average or with data quality flags. We additionally excluded FFIs obtained when the spacecraft’s pointing was compromised, when TESS was affected by an instrument anomaly, or when significant scattered light was present. 

We converted the TESS counts into fluxes using the instrumental zero point of 20.44 electrons per second in the FFIs \citep{vanderspek18}. For the TESS data during the flares, we calibrated them to concurrent ATLAS $o$-band data, as the TESS effective wavelength of $\sim$$7500$~\AA\ is similar to the $\sim$$6750$~\AA\ effective wavelength of the ATLAS $o$-band. This was done by shifting the TESS light curve by a linear offset to match the ATLAS photometry and then scaling by a multiplicative factor to correct the light curve shape. To ensure a high S/N per epoch, we stacked the TESS data in 2-hour bins. The TESS photometry in all sectors is shown with brown octagons in Figure \ref{fig:long_lc}. The TESS data during the flares is shown in Figure \ref{fig:opt_lc} and presented in Table \ref{tab:phot}. We caution that the TESS sectors during the flares of ASASSN-22ci were affected by high amounts of scattered light and are likely affected by systematics, particularly near the ends of the sectors and the data downlink gaps.

From the pre-flare Sector 22 data, we obtained a strong limit on any pre-flare variability. The root-mean-squared (RMS) scatter is 8 $\mu$Jy and decreases to 6 $\mu$Jy when subtracting the mean flux uncertainty in quadrature. Treating this RMS scatter as a rough estimate of the pre-existing AGN variability, this flux corresponds to $\lambda \textrm{L}_{\lambda} = 5 \times 10^{40}$ erg s$^{-1}$ at the distance of ASASSN-22ci. For a conservative AGN fractional variability of 1\% and the SMBH mass from Section \ref{sec:archival}, this constrains the Eddington ratio to be $\lesssim 0.02$, suggesting ASASSN-22ci does not host a strong AGN, consistent with the pre-flare X-ray constraints.

\begin{deluxetable}{ccc}
\tablewidth{240pt}
\tabletypesize{\footnotesize}
\tablecaption{Synthetic \swift UVOT Host-Galaxy Magnitudes}
\tablehead{
\colhead{Filter} &
\colhead{AB Magnitude} &
\colhead{Magnitude Uncertainty}}
\startdata
$UVW2$ & 21.21 & 0.15 \\
$UVM2$ & 21.12 & 0.12 \\
$UVW1$ & 19.97 & 0.08 \\
$U$ & 18.23 & 0.09 \\
$B$ & 16.91 & 0.07 \\
$V$ & 16.15 & 0.05 \\
\enddata 
\tablecomments{Synthetic magnitudes of the host galaxy in the \swift UVOT filters, used to subtract the host component from the UVOT photometry. The uncertainties are estimated from the standard deviation of 10,000 Monte Carlo iterations of the host SED fit. All magnitudes are in the AB system.} 
\label{tab:swift_host_mags} 
\end{deluxetable}

\subsection{\swift Observations}

We analyzed seventy-one Neil Gehrels Swift Gamma-ray Burst Mission (\swiftn; \citealt{gehrels04}) target-of-opportunity (ToO) observations taken between 2022 February 27 and 2024 May 20 (Swift target ID 15026, PIs: Hinkle, Holoien, Margutti, Lin). Each of these observations obtained data with the UltraViolet and Optical Telescope (UVOT; \citealt{roming05}) and the X-ray Telescope (XRT; \citealt{burrows05}) aboard \swiftn.

\subsubsection{UVOT Observations}

For most \swift epochs, ASASSN-22ci was observed with all six of the UVOT filters typically used for TDE follow-up: $V$ (5425.3 \AA), $B$ (4349.6 \AA), $U$ (3467.1 \AA), $UVW1$ (2580.8 \AA), $UVM2$ (2246.4 \AA), and $UVW2$ (2054.6 \AA), where the wavelengths are the pivot wavelengths from the SVO Filter Profile Service \citep{rodrigo12}. We first summed images within the same observation using the HEASoft \texttt{uvotimsum} package. To compute the photometry, we used \texttt{uvotsource} with a 10\farcs{0} radius region centered on the TDE position and a source-free background region with a radius of 50\farcs{0}. We use a 10\farcs{0} radius aperture to minimize the effects of the spacecraft pointing jitter that was present during the second flare. 

There is no UVOT data available for the host galaxy prior to ASASSN-22ci. We therefore estimated the flux of the host in the \swift UVOT bands by computing synthetic photometry from the best-fit \textsc{Fast} host SED. These fluxes, shown in Table \ref{tab:swift_host_mags}, were then subtracted from the measured \swift photometry to yield the transient-only photometry. The UV photometry of ASASSN-22ci from \swift is shown in Figure \ref{fig:opt_lc} and given in Table \ref{tab:phot}. While ASASSN-22ci is largely undetected between the flares, there are two weak ($\sim$$3-4\sigma$) detections in the $UVW2$ band and one $\sim 3\sigma$ detection in the $B$ band. The other UV bands have only upper limits between the flares, although they too generally show weak excess emission at the $\sim$$2\sigma$ level.

\begin{deluxetable}{ccccc}
\tablewidth{240pt}
\tabletypesize{\footnotesize}
\tablecaption{Host-Subtracted Photometry of ASASSN-22ci}
\tablehead{
\colhead{MJD} &
\colhead{AB Mag} &
\colhead{Uncertainty} &
\colhead{Filter} &
\colhead{Source}}
\startdata
59637.2870 & 15.91 & 0.04 & UVW2 & Swift \\
59640.4800 & 16.01 & 0.04 & UVW2 & Swift \\
59642.4010 & 16.22 & 0.04 & UVW2 & Swift \\
\dots  & \dots  & \dots & \dots  & \dots  \\
60394.0848 & 19.28 & 0.08 & TESS & TESS \\
60394.1692 & 19.30 & 0.08 & TESS & TESS \\
60394.2456 & 19.04 & 0.06 & TESS & TESS \\
\enddata 
\tablecomments{Host-subtracted and extinction-corrected magnitudes and 3$\sigma$ upper limits for our follow-up photometry. All magnitudes are presented AB system and a value of 99.99 in the uncertainty column indicates an upper limit. Only a small section of the table is displayed here to show the format. The full table can be found online as an ancillary file.} 
\label{tab:phot} 
\end{deluxetable}

\subsubsection{XRT Observations}\label{sec:xrt}

\begin{figure}
\centering
 \includegraphics[width=0.36\textwidth]{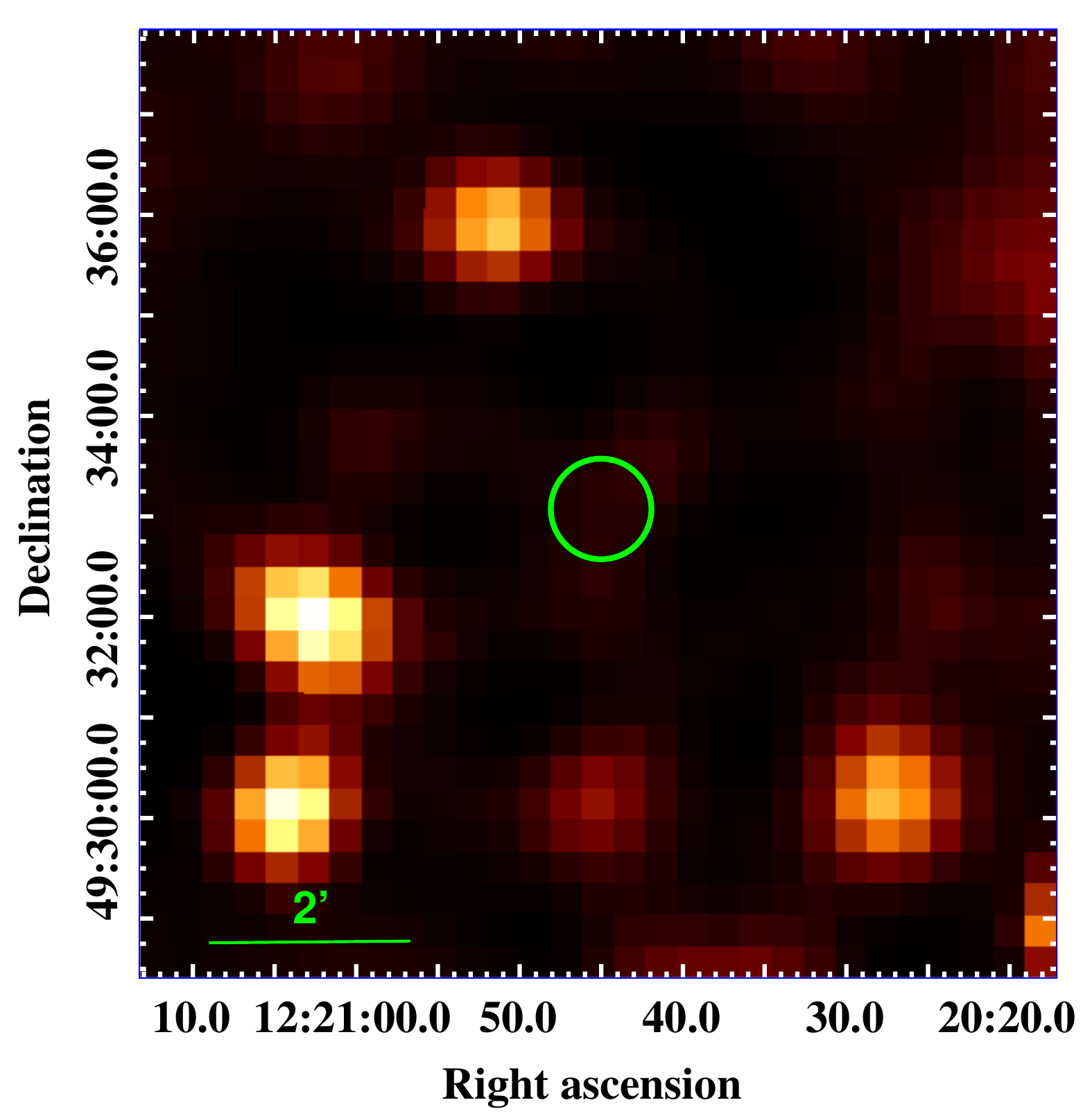}
  \includegraphics[width=0.36\textwidth]{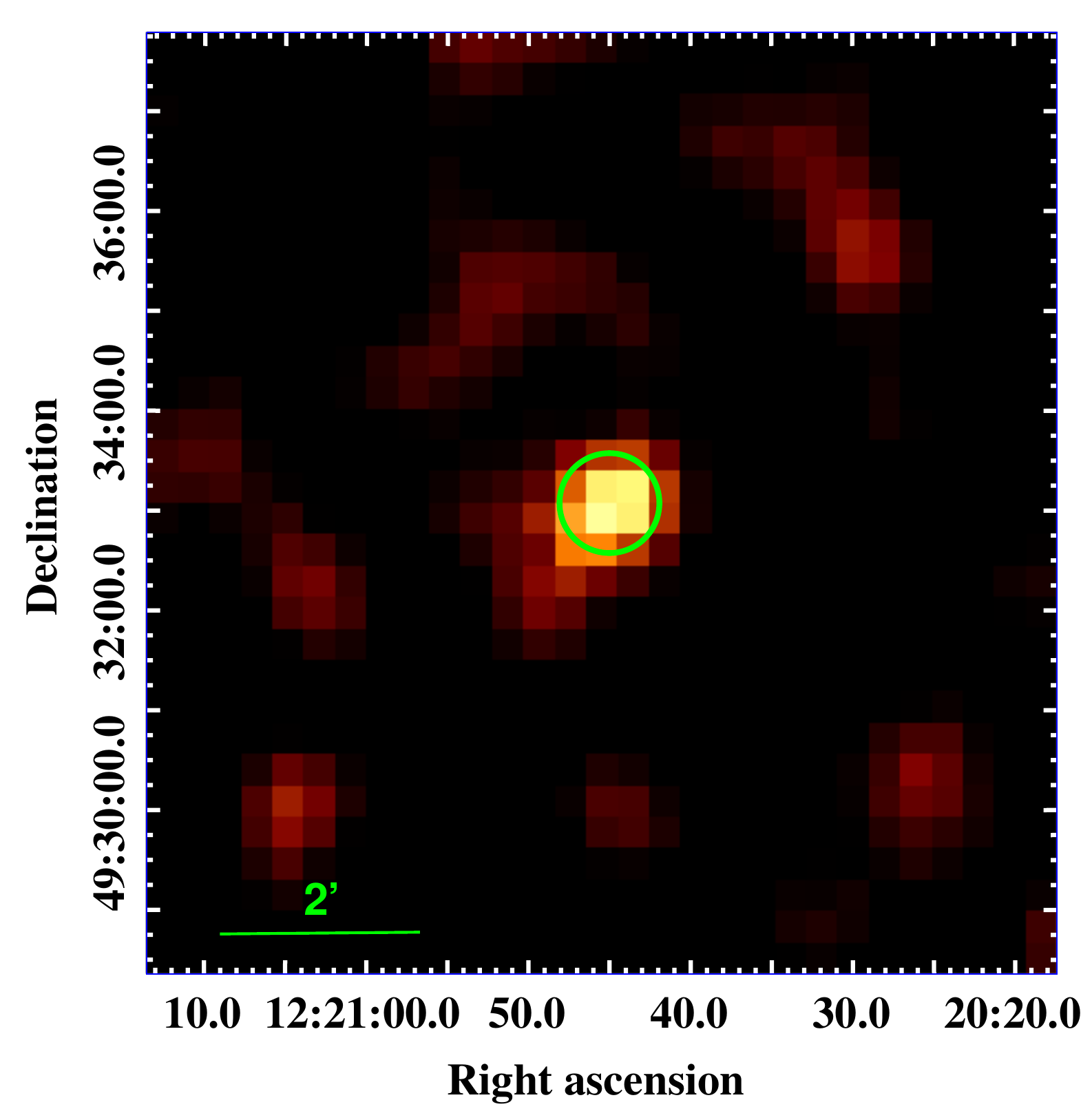}
   \includegraphics[width=0.36\textwidth]{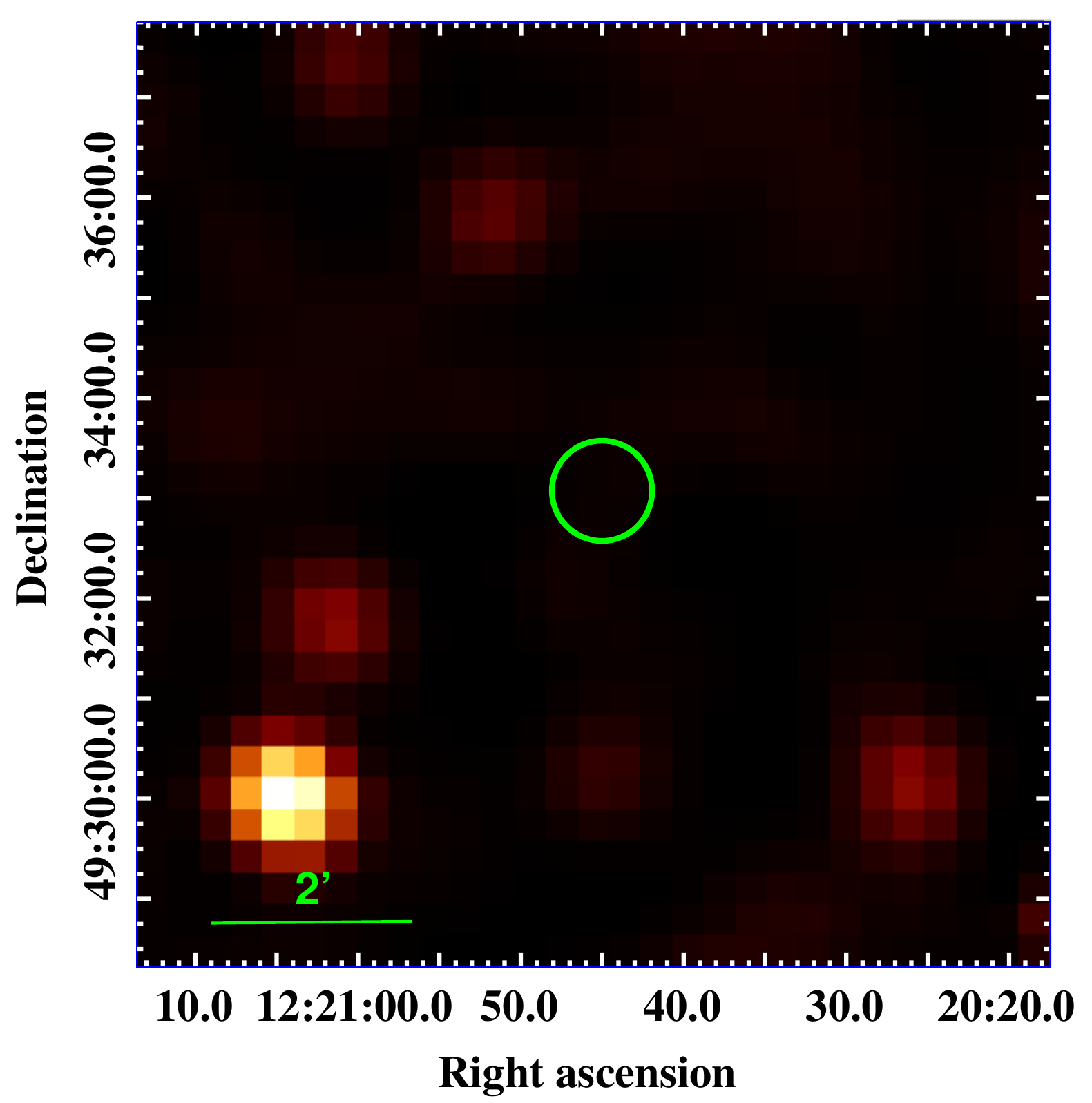}
 \caption{Broad-band (0.3-10.0 keV) merged \textit{Swift} XRT observations of (top) the first flare, (middle) between the two flares, and (bottom) during the second flare. The green circle marks the location of ASASSN-22ci. The exposure time of the merged dataset associated with Flare 1 (top) is 78.7 ks, between the flares (middle) is 2.3 ks and Flare 2 (bottom) is 60.5 ks.}
 \label{fig:xray_images}
\end{figure}

Concurrent with the \swift UVOT observations, we also obtained \swift X-Ray Telescope (XRT) photon-counting observations of ASASSN-22ci. We reprocessed all observations from level one XRT data using the package \textsc{xrtpipeline} version 0.13.7 and applied standard filter and screening criteria\footnote{\url{http://swift.gsfc.nasa.gov/analysis/xrt_swguide_v1_2.pdf}} with the most recent calibration files. Using a source region with a radius of 47\farcs{0} centered on the position of ASASSN-22ci and a source-free background region with a radius of 150\farcs{0} centered at $(\alpha,\delta)=$ (12:20:42.1406,$+$49:40:22.740), we found no significant X-ray emission associated with individual observations of both flares. However, we found significant X-ray emission associated with ObsIDs: 00015026043 and 00015026044, which were taken 374 and 377 days after peak brightness (MJD 60009.71 and 60012.68, respectively), well after the first flare had faded and before the second flare began. Significant, although weak, emission in the Swift $UVW2$ filter was also detected at these epochs (see Figure \ref{fig:opt_lc}).  

To increase the S/N of our observations, we grouped the individual \swift observations into three time bins using \textsc{xselect} version 2.5b. These bins were named Flare 1 (F1: MJDs 59636 to 59864), Between Flares (BF: MJDs 60009 to 60013), and Flare 2 (F2: MJDs 60235 to 60450). Figure \ref{fig:xray_images} shows the broadband (0.3-10.0 keV) X-ray images created from the three groups. During the first flare (group F1) and second flare (group F2), there is no detected X-ray emission, with 0.3-10.0 keV 3$\sigma$ upper-limits of $<0.002$ counts/sec and $<0.001$ counts/sec, respectively. However, in group BF, between the two  UV/optical flares, ASASSN-22ci becomes X-ray bright, with significant ($>4\sigma$) X-ray emission, which is visible in the middle panel of Figure \ref{fig:xray_images}. This emission has a 0.3-10.0 keV count rate of 0.004$\pm$0.001 counts/sec. Assuming an absorbed blackbody with a column density of $N_{H}=1.94\times10^{20}$ cm$^{-2}$ and a temperature of 0.042 keV (see Section \ref{sec:xray} for more details), we derive an X-ray flux for group BF of $(3.5\pm1.0) \times10^{41}$ erg s$^{-1}$. For group F1 and F2, we derive 3$\sigma$ upper-limits to the luminosity of $<1.5\times10^{41}$ erg s$^{-1}$ and $<9.5\times10^{40}$ erg s$^{-1}$, respectively.
 
We extracted a low-count spectrum from the BF data using the task \textsc{xrtproducts} and the source and background regions from above. We made ancillary response files (ARF) with the task \textsc{xrtmkarf} and the individual exposure maps generated by \textsc{xrtpipeline}. We made use of the ready-made response matrix files (RMFs). Due to the low count rate, we grouped this spectrum to have a minimum of 1 count per energy bin using the \textsc{ftools} command \textsc{grppha}, and use the Bayesian X-ray Analysis \citep[BXA;][]{2014A&A...564A.125B} which connects the nested sampling algorithm UltraNest \citep{2021JOSS....6.3001B} with the fitting environment \textsc{xspec} \citep{1996ASPC..101...17A}.

\begin{figure*}
\centering
 \includegraphics[width=1.0\textwidth]{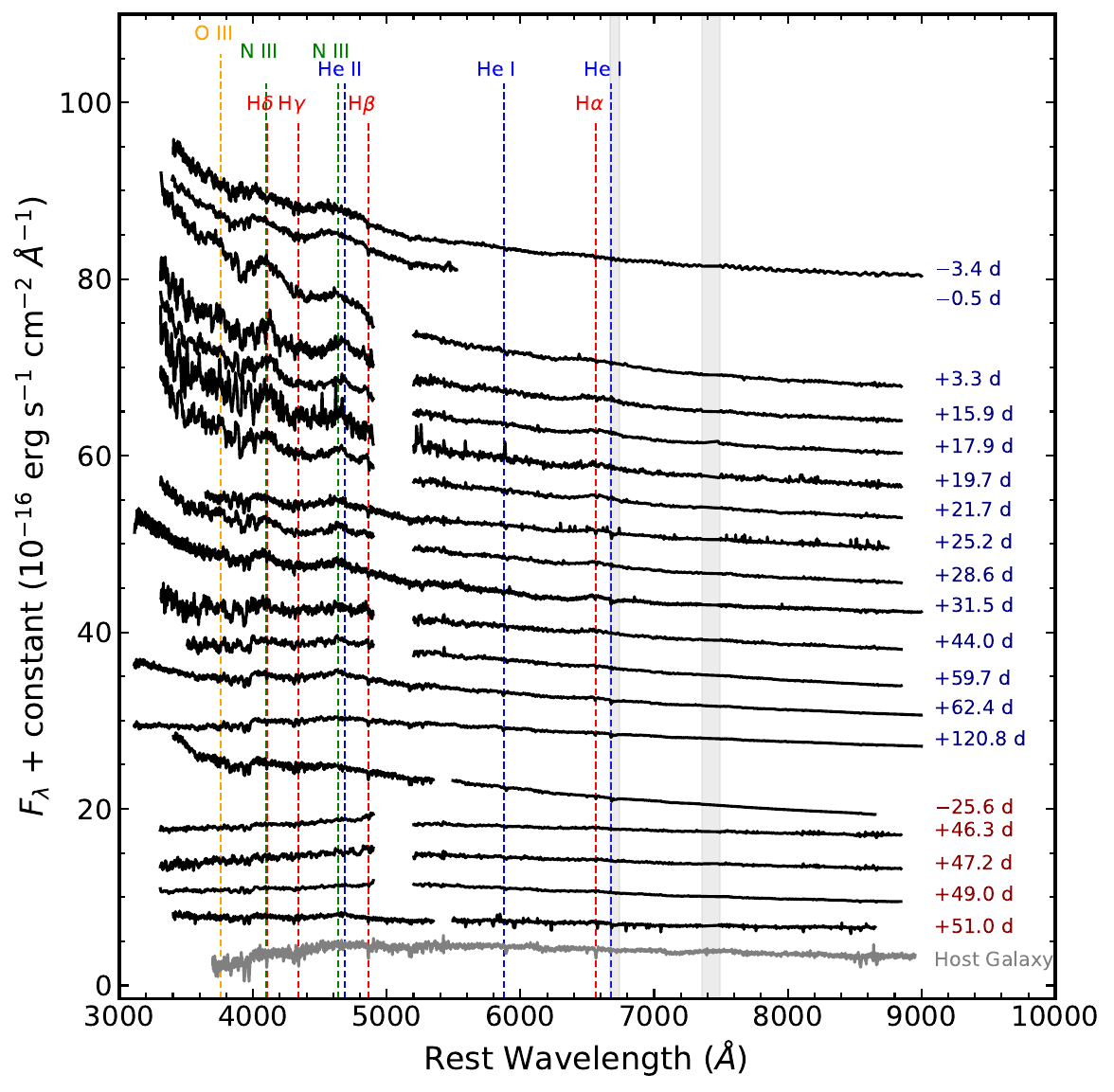}
 \caption{Optical spectroscopic evolution of ASASSN-22ci with the phases during the first flare colored in navy and those during the second flare marked in red. The spectra are calibrated using the optical photometry presented in Fig.\ \ref{fig:opt_lc}. The vertical gray bands mark atmospheric telluric features. The regions corresponding to the dichroic crossovers in the SNIFS and KCWI spectra have been excised. An archival host spectrum from SDSS is shown at the bottom of the figure in gray. The vertical lines mark spectral features commonly seen in TDEs. These are the Balmer lines of hydrogen in red, helium lines in blue, nitrogen lines in green, and oxygen lines in orange.}
 \label{fig:opt_spec}
\end{figure*}

\subsection{Spectroscopic Observations}

We obtained follow-up spectra of ASASSN-22ci from several sources. These include the SuperNova Integral Field Spectrograph \citep[SNIFS;][]{lantz04} on the 2.2-m University of Hawaii telescope (UH2.2) as part of the Spectroscopic Classification of Astronomical Transients \citep[SCAT;][]{tucker22} survey, the Low-Resolution Imaging Spectrometer \citep[LRIS;][]{oke95} on the 10-m Keck I telescope, the Multi-Object Double Spectrographs \citep[MODS;][]{pogge10} on the Large Binocular Telescope \citep[LBT; ][]{hill06}, and the Keck Cosmic Web Imager \citep[KCWI;][]{morrissey18} on the 10-m Keck II telescope. Finally, we obtained the public classification spectrum taken with FLOYDS from TNS \citep{arcavi22}.

We reduced the SNIFS spectra with the custom SCAT pipeline \citep{tucker22}, the LRIS and KCWI spectra with \textsc{PypeIt} \citep{pypeit_prochaska}, and the MODS spectrum with standard IRAF procedures. We refined the initial flux calibration of our spectra, typically from standard star spectra obtained on the same night, by mangling the spectra to match our follow-up photometry. To avoid negative fluxes, especially on the reddest end, we added 10\% of the host-galaxy flux in quadrature to the transient photometry used to scale the spectra.

These optical spectra are shown in Figure \ref{fig:opt_spec}. The earliest spectra, shown at the top of the figure, exhibit a hot, blue continuum with broad emission lines consistent with \ion{He}{2} and \ion{N}{3}. A weak H$\alpha$ line is present and grows in time with the other emission features. After peak, the spectra fade and become less blue, although retaining a UV excess through the last spectrum of the first flare. The spectra in the second flare show nearly identical features, although they are generally weaker as the second flare was fainter. We mark the locations of several common TDE emission lines with vertical dashed lines in Fig.\ \ref{fig:opt_lc}.

\begin{figure*}
\centering
 \includegraphics[width=1\textwidth]{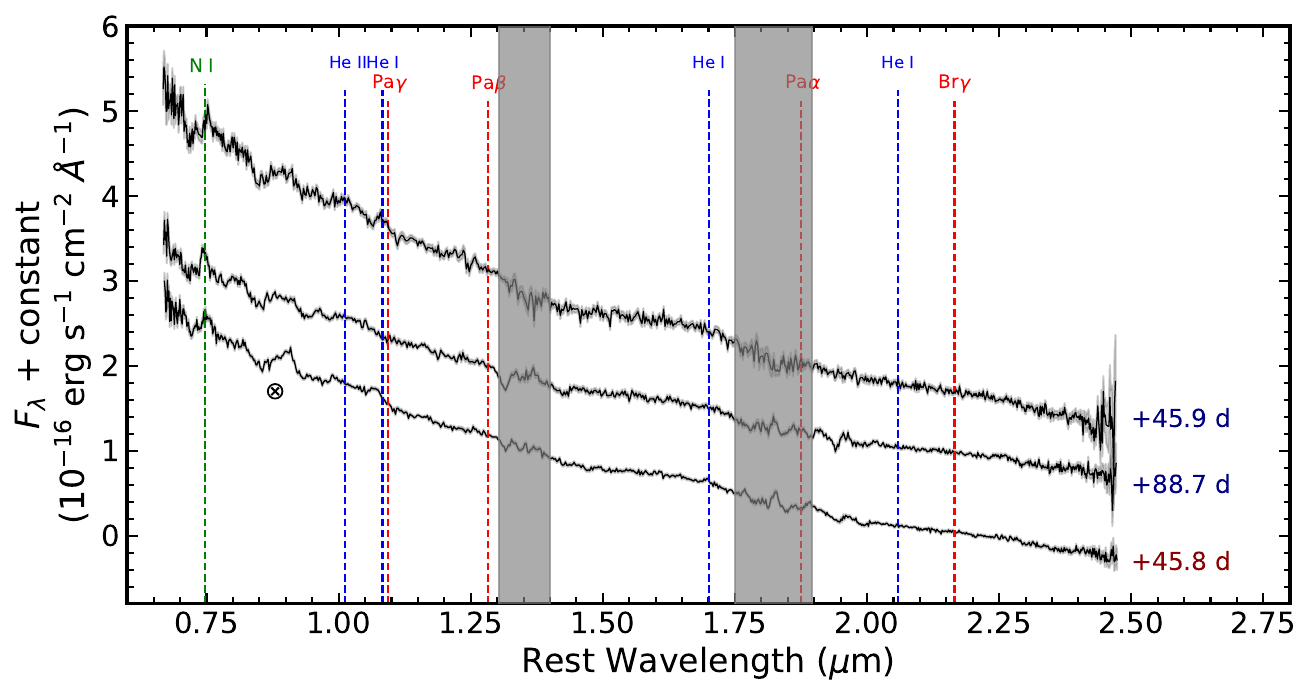}
 \caption{Near-infrared spectra of ASASSN-22ci taken with IRTF/SpeX in Prism mode. The rest-frame phases of the spectra from the peak of the first (second) flare are given at the right in navy (red). The vertical gray bands mark strong atmospheric telluric features and the $\otimes$ marks a likely instrumental feature. The vertical lines mark spectral features from hydrogen in red, helium in blue, and nitrogen in green.}
 \label{fig:nir_spec}
\end{figure*}

In addition to the optical spectra of ASASSN-22ci, we obtained three near-infrared (NIR) spectra with SpeX \citep{rayner03} on the NASA Infrared Telescope Facility (IRTF), two during the first flare and one during the second. These data were obtained in Prism mode, giving us coverage over roughly the $zYJHK$ bands at low resolution ($R \approx 80$). These spectra were reduced and calibrated using SpeXtool \citep{cushing04}. A telluric standard taken on same night and at similar airmass was used to remove atmospheric features and for the flux calibration. These spectra are shown in Figure \ref{fig:nir_spec}.

\section{Analysis} \label{sec:analysis}

In this section, we present our analysis of the UV/optical light curves, optical/NIR spectroscopy, UV/optical spectral energy distribution, and X-rays from ASASSN-22ci.

\subsection{Light Curve} \label{sec:lc}

ASASSN-22ci has TESS coverage in Sectors 48 and 49, very close to the beginning of the flare. We can use this TESS data to constrain the parameters of the early-time rise. We fit the TESS flux with
\begin{equation}
f(t) = \frac{h}{(1+z)^2} \bigg(\frac{t - t_1}{1+z}\bigg)^{\alpha_1 \big(1 + \alpha_2 \times (t - t_1)/(1+z)\big)} + f_0
\end{equation}
for $t > t_1$ and $f(t) = f_0$ for $t < t_1$ \citep{vallely21}. Here $z$ is the source redshift, $h$ is a flux scaling, $t_1$ is the time of first light, $a_1$ is the initial power-law slope, $a_2$ is the second power-law slope, and $f_0$ is an overall flux constant. The inclusion of a second power-law index $\alpha_2$ allows for the curvature of the light curve and minimizes biases in the early-time power-law index $\alpha_1$. We fit only the Sector 48 TESS data, MJDs 59609.9 through 59635.5, as this is before the peak of the TESS flux and allows us to avoid inter-sector flux calibration issues.

\begin{figure}
\centering
 \includegraphics[width=0.49\textwidth]{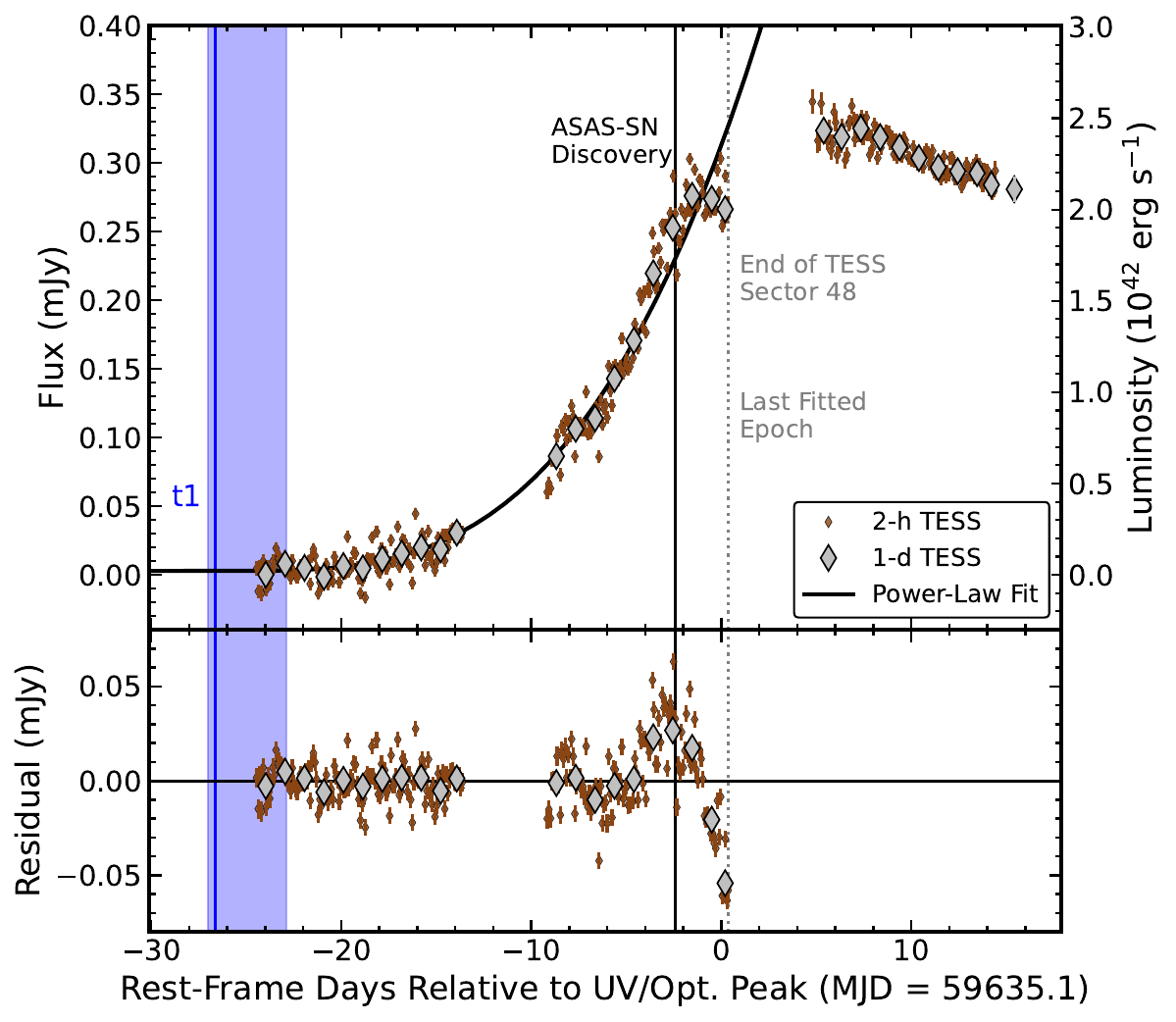}
 \caption{\textit{Top panel}: The 2-h binned (red) and 1-d binned (silver) TESS data and best-fitting two-component power-law model in black. The power-law fit yields a time of first light of $t_1 \textrm{(MJD)} = 59607.7^{+3.7}_{-0.4}$ and an initial power-law index of $\alpha_1 = 3.5^{+0.2}_{-0.8}$. The vertical blue line indicates the time of first light $t_1$ with the shading representing the uncertainty on $t_1$. The vertical black line is the time of ASAS-SN discovery and the vertical dotted line is the last epoch included in the power-law fit. \textit{Bottom panel}: Residuals between the data and the best-fit power-law model. The TESS light curves often exhibit systematic trends near the beginning and end of a sector, so the downturn shortly after the ASAS-SN discovery epoch may be due to systematics.}
 \label{fig:TESS_rise}
\end{figure}

A Markov Chain Monte Carlo (MCMC) fit using this model yields best-fit parameters of $f_0 = 2.8^{+1.9}_{-0.6} \  \mu\text{Jy}, \ h = 4.6^{+70}_{-1.6} \times 10^{-6} \ \text{Jy}, \ t_1 \text{(MJD)} = 59607.7^{+3.7}_{-0.4}, \ \alpha_1 = 3.5^{+0.2}_{-0.8}$, \ \text{and} \ $\alpha_2 = -5.4^{+5.0}_{-13} \times 10^{-4}$. The best-fit values are the maximum likelihood model and the uncertainties represent the 90\% confidence interval. This fit, along with the full TESS light curve for the first flare are shown in Figure \ref{fig:TESS_rise}. We also fit a single-power law model ($\alpha_2 = 0$) through data up to 40\% of the peak flux  \citep{vallely21} and find consistent results for $\alpha_1$.
 
ASASSN-22ci has a steeper initial rise slope than any previous TDE with early-time rise constraints \citep[e.g.,][]{holoien19c, hinkle21a, nicholl20, payne21, hoogendam24}. The rise time of $\approx$27 days is shorter than the rise time measured for ASASSN-19bt \citep{holoien19c}, similar to the rise times for the TDE ASASSN-19dj \citep{hinkle21a} and AT2019qiz \citep{nicholl20}, and longer than the rise times for ASASSN-23bd \citep{hoogendam24}, ASASSN-14ko \citep{payne21, payne23}, and AT2020neh \citep{angus22}. 

\begin{figure*}
\centering
\includegraphics[width=0.48\textwidth]{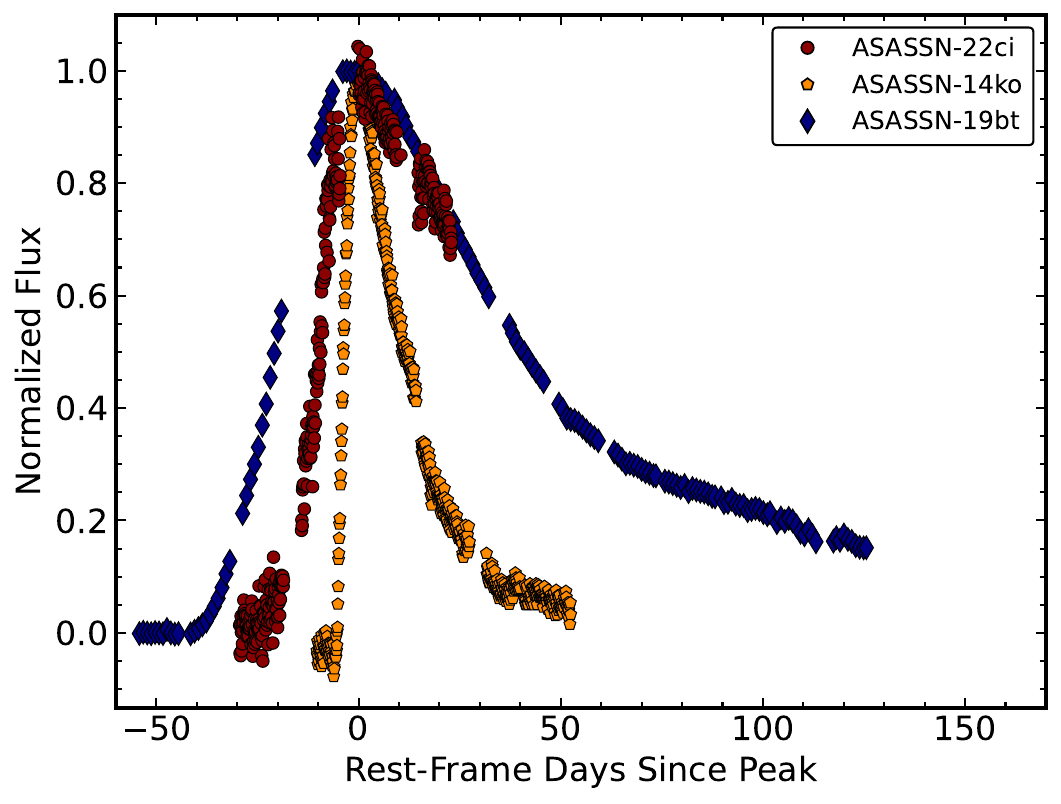}
 \includegraphics[width=0.48\textwidth]{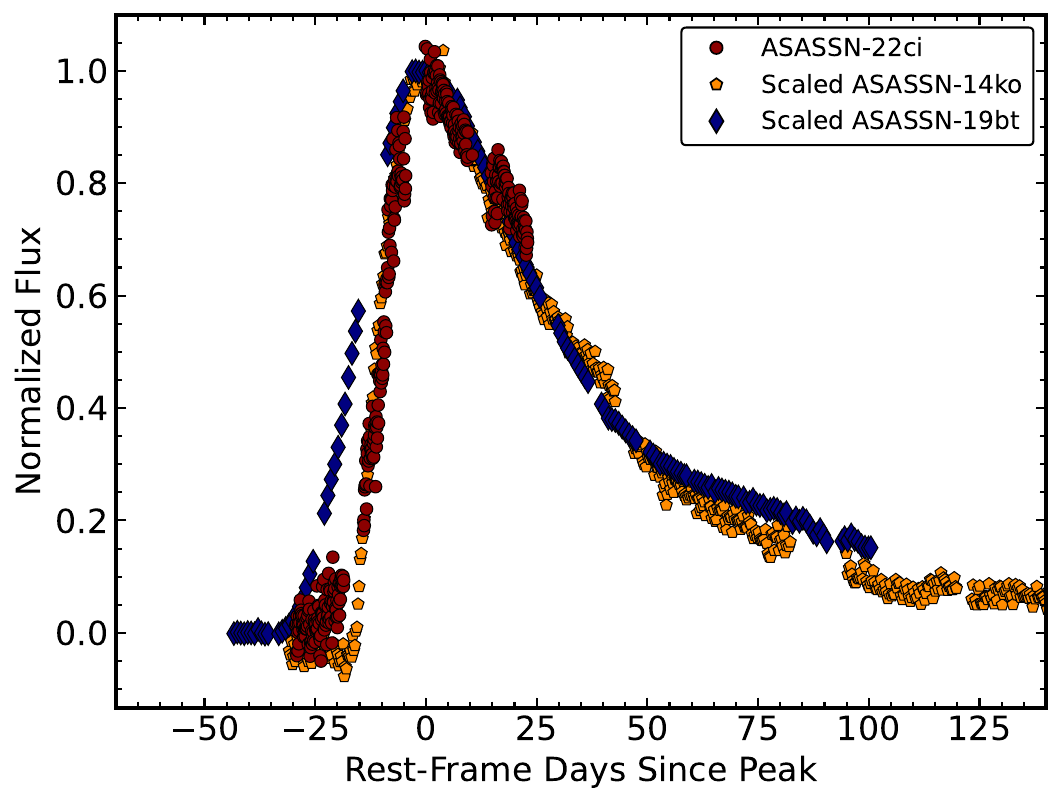}
 \caption{TESS light curves of ASASSN-22ci (red circles), ASASSN-14ko (orange pentagons), and ASASSN-19bt (navy diamonds) each binned at 2 hours. The left panel shows the rest-frame TESS light curves of each TDE. The right panel shows the ASASSN-22ci light curve, the ASASSN-14ko light curve expanded 3$\times$ in time, and the ASASSN-19bt light curve compressed by a factor of 0.8 to better align the post-peak declines.}
 \label{fig:TESS_comp}
\end{figure*}

Figure \ref{fig:TESS_comp} compares the high cadence and high S/N TESS light curves available for ASASSN-22ci, ASASSN-14ko, and ASASSN-19bt. The flares exhibit a range of rise and decay timescales, with ASASSN-22ci intermediate to the shorter timescale ASASSN-14ko flares and the longer timescale flare of ASASSN-19bt. However, the flares show some similarities when we expand the timescale of ASASSN-14ko by a factor of 3 and compress the timescale for ASASSN-19bt by a factor of 0.8 to align the declines. ASASSN-14ko rises the fastest but at a shallower slope \citep{payne21} than either ASASSN-22ci or ASASSN-19bt. While the rise of ASASSN-22ci begins earlier than that of ASASSN-14ko, by the time each flare reaches $\sim$30\% of its peak flux, the shapes of the flares are nearly identical. Like almost all TDEs, none of the flares exhibits any significant variability near peak.

\subsection{Spectra} \label{sec:spectra}

The spectra of ASASSN-22ci taken before and near peak light exhibit the canonical blue continuum and broad emission lines of TDEs. A blue/near-UV excess relative to the host-galaxy spectrum continues through peak and for at least two months thereafter. We see strong lines of \ion{N}{3}, \ion{O}{3}, and H, possibly with \ion{He}{2} making this event a member of the TDE H+He class, and the N+O subclass, in particular, \citep{leloudas19, vanvelzen20b}. The spectra of ASASSN-22ci are similar to other TDEs of the same spectroscopic class such as iPTF-15af \citep{blagorodnova18}, iPTF16axa \citep{hung17}, and ASASSN-18pg \citep{leloudas19, holoien20}. Compared to many of these events, the Balmer emission of ASASSN-22ci is quite weak, with no significant H$\beta$ or higher-order Balmer lines seen at any point in the source evolution.

To quantify the spectroscopic evolution we measured the line widths and luminosities of the \ion{N}{3} $\lambda 4100$ line, \ion{N}{3} $\lambda 4640$/\ion{He}{2} $\lambda 4686$ line blend, and H$\alpha$ emission line. To do this, we first estimate and remove the continuum emission to isolate the emission lines. We estimated the continuum using the AstroPy method \textsc{fit\_generic\_continuum}. For the \ion{N}{3} $\lambda 4100$ and \ion{N}{3} $\lambda 4640$/\ion{He}{2} $\lambda 4686$ features, we estimated the continuum using the wavelength ranges of $3000\mathrm{\AA}\leq\lambda\leq3900\mathrm{\AA}$ and $5000\mathrm{\AA}\leq\lambda\leq5500\mathrm{\AA}$. As the KCWI spectra do not have continuous coverage over the dichroic crossover region, it was necessary to add an additional red continuum region. For the H$\alpha$ line, we used the wavelength ranges of $3400\mathrm{\AA}\leq\lambda\leq3600\mathrm{\AA}$, $5000\mathrm{\AA}\leq\lambda\leq6200\mathrm{\AA}$, and $7100\mathrm{\AA}\leq\lambda\leq8500\mathrm{\AA}$. For the spectra taken after MJD 59699, we excluded the bluest wavelength region from our continuum fits as these resulted in better fits to the continuum.

\begin{figure}
\centering
 \includegraphics[width=0.49\textwidth]{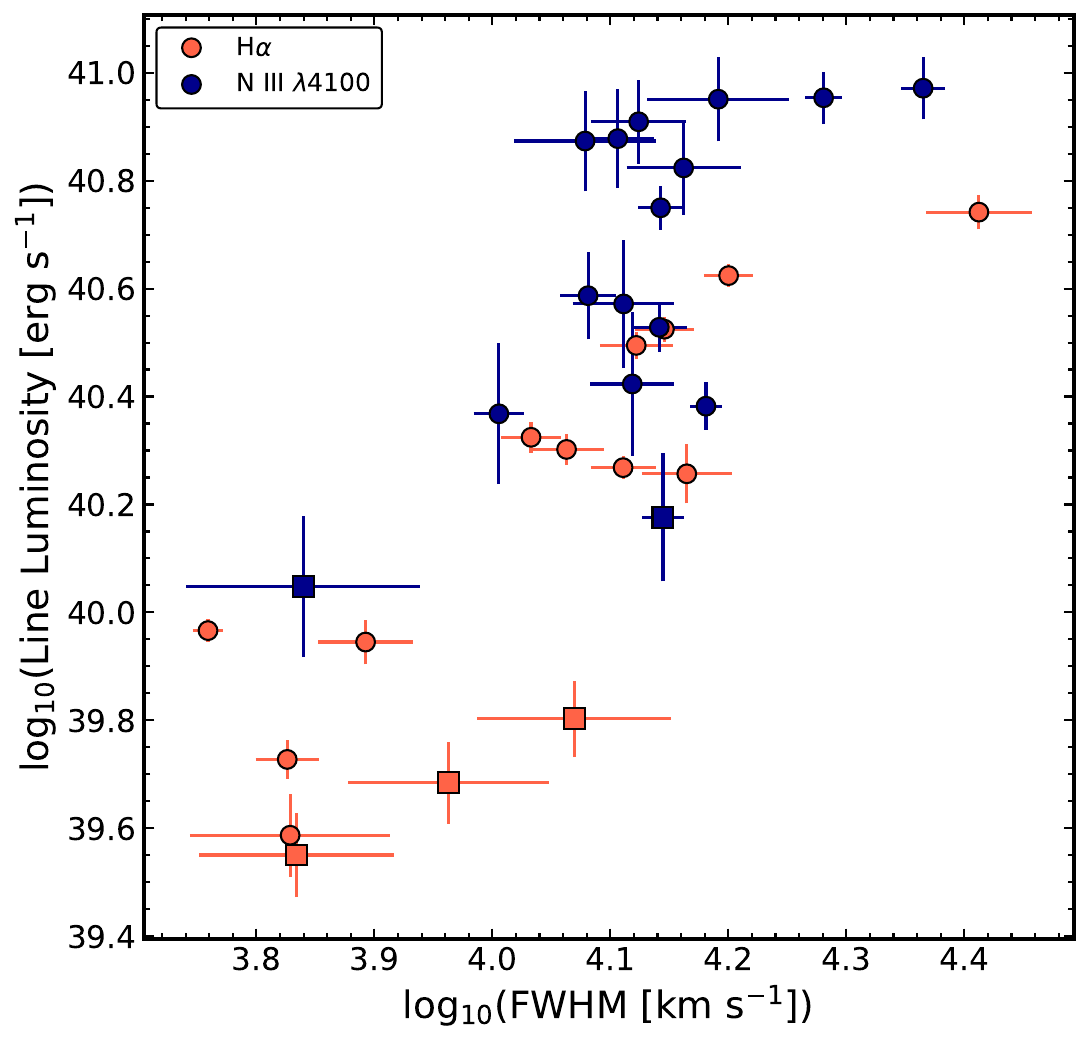}
 \caption{H$\alpha$ (red points) and \ion{N}{3} $\lambda4100$ (blue points) emission line luminosity as compared to the FWHM of the line. Spectra taken during the first (second) flare are plotted with circles (squares). Each line shows a positive correlation between luminosity and FWHM.}
 \label{fig:22ci_emission_lines}
\end{figure}

After subtracting the fitted continuum from the spectra, we modeled the emission lines as Gaussians. We fit the blue portion of the spectrum with two Gaussian components, nominally one each for the \ion{N}{3} $\lambda 4100$ line and \ion{N}{3}$\lambda 4640$/\ion{He}{2} $\lambda 4686$ line blend, finding good fits. We were unable to fit the three SNIFS spectra during the second flare as we could not adequately fit the continuum due to issues with the dichroic crossover region. We fit H$\alpha$ as a single Gaussian. Spectra after the $+120.8$ day LRIS spectrum had very faint H$\alpha$ features, and we could not find reliable fits. The FWHM and line luminosity for the isolated \ion{N}{3} $\lambda 4100$ and H$\alpha$ lines are shown in Figure \ref{fig:22ci_emission_lines}. The \ion{N}{3} luminosities are roughly an order of magnitude higher than for H$\alpha$, although they have a similar median FWHM.

The FWHM and luminosity of the emission lines seen for ASASSN-22ci are positively correlated, similar to other TDEs \citep[e.g.,][]{holoien14b, hinkle21a}. The FHWM and luminosity of the \ion{N}{3}$ \lambda 4100$ line fits are moderately correlated, with Kendall $\tau = 0.38$ and a p-value of 0.04 when considering both flares. We find a stronger correlation for H$\alpha$, with Kendall $\tau = 0.58$ and a p-value of 0.002. These correlations are driven primarily by data for the first flare, for which the correlations are similarly strong.

The NIR spectra of ASASSN-22ci have weaker features than the optical, likely due to stronger dilution by the host-galaxy flux in the NIR. Nevertheless, several features are present and appear to evolve, including likely \ion{He}{1} $1.083 \mu$m, and possible \ion{N}{1} $\lambda0.7468 \mu$m and \ion{He}{2} $\lambda1.012 \mu$m. The \ion{He}{1} $\lambda1.083 \mu$m line in particular shows strong evolution. It is relatively strong in the $+46$ day spectra in the two flares, but absent in the $+88.7$ day spectrum from the first flare. 

NIR spectral flattening is predicted by some reprocessing models \citep{roth16, lu20}. While the NIR spectra of ASASSN-22ci are flatter than the Rayleigh-Jeans tail of a blackbody, the strong host contribution makes it difficult to determine if this is consistent with a reprocessing scenario. There are no strong lines in the optical or NIR consistent with coronal lines \citep[e.g.,][]{wang11, yang13, hinkle24a}. This suggests either a lack of the soft X-ray and extreme UV photons needed to produce the necessary ions or a lack of dense \citep[$n_e \sim 10^7$ cm$^{-3}$;][]{komossa09, wang12, wang24} gas near the central SMBH.

Using the KCWI data cube obtained on MJD 60328.6, we searched for spatially extended emission, as was recently found for several post-starburst galaxies \citep{french23} and TDE hosts \citep{tucker21, wevers24}. We created an [\ion{O}{3}] $\lambda 5007$ image by summing the slices within 300 km s$^{-1}$ of the line center and subtracted a continuum image constructed from slices between 500 and 1000 km s$^{-1}$ from the line on either side. We find no evidence for extended [\ion{O}{3}] emission at any position within the KCWI cube. At the distance of ASASSN-22ci, the field of view of the KCWI images corresponds to approximately 10 kpc, larger than the projected distance to the extended emission-line regions seen for Mrk 950, the host galaxy of the TDE iPTF-16fnl \citep{wevers24}. The 0\farcs{7} spatial resolution of our KCWI cube corresponds to a physical distance of 400 pc at the host galaxy of ASASSN-22ci, the smallest scale over which we can constrain the presence of extended emission. 

From emission line diagnostics, WISE MIR measurements of the host-galaxy SED, TESS photometric constraints, and X-ray upper limits we have already ruled out the existence of a luminous AGN in the host galaxy of ASASSN-22ci. The lack of extended emission [\ion{O}{3}] between $\approx$$0.4 \mbox{ -- } 10$ kpc similarly suggests that this galaxy did not host a strong AGN between $\approx$1300 and $\approx$30,000  years ago. Since the typical AGN lifetime is longer than either of these constraints \citep[e.g.][]{marconi04}, it is likely that this galaxy has not hosted a strong AGN within the past 
$\approx$30,000 years.

\subsection{UV/Optical Spectral Energy Distribution} \label{sec:sed}

\begin{figure*}
\centering
 \includegraphics[width=0.99\textwidth]{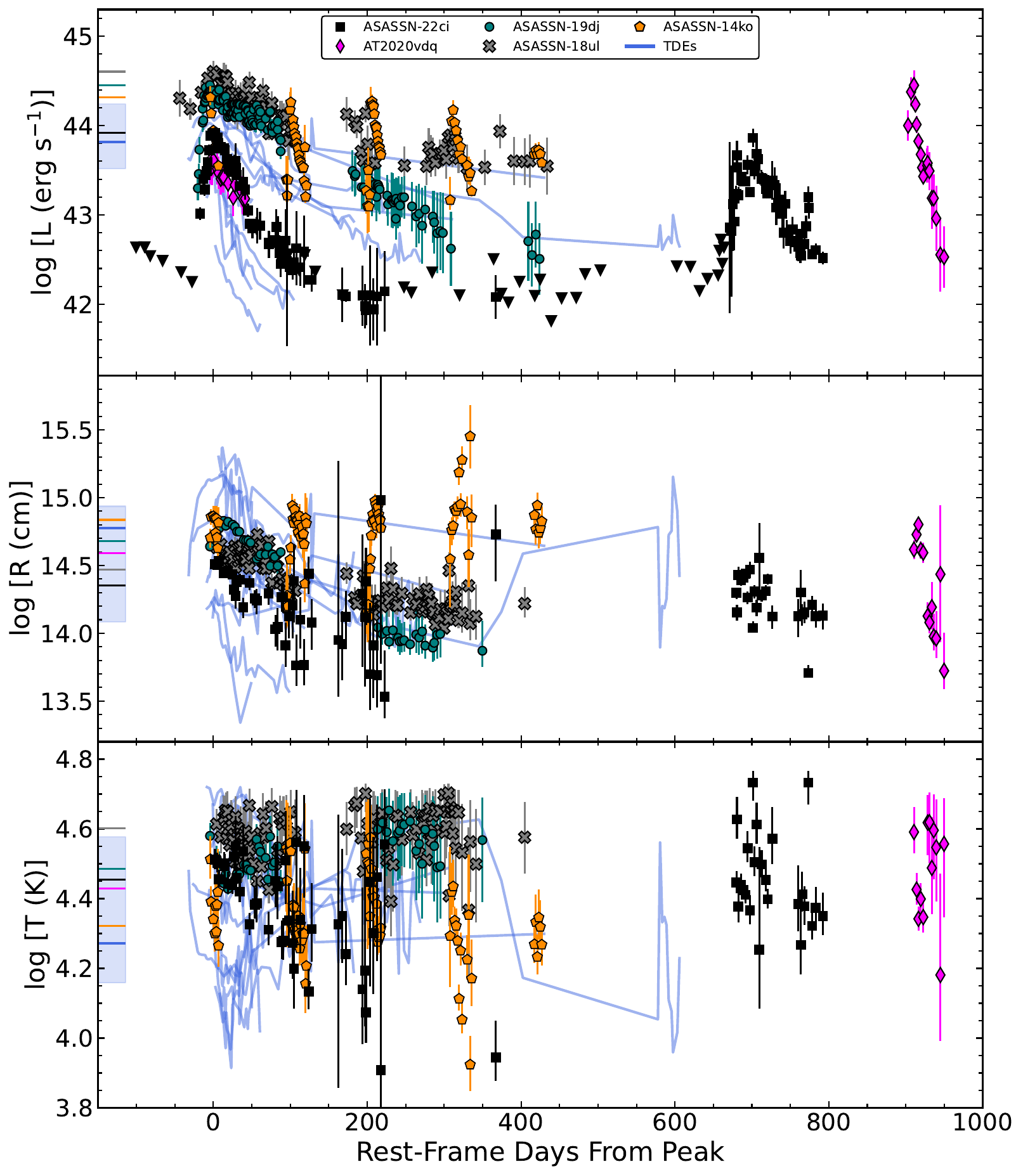}
 \caption{Temporal evolution of the bolometric UV/optical luminosity (top panel), effective radius (middle panel), and effective temperature (bottom panel) for ASASSN-22ci (black squares). Also shown are the comparison multiple/repeating TDE candidates ASASSN-14ko (orange pentagons), ASASSN-18ul (gray crosses), ASASSN-19dj (teal circles), and AT2020vdq (magenta diamonds). AT2020vdq has no UV data for the first flare and therefore we cannot reliably estimate a radius and temperature evolution. We only show the flares for ASASSN-14ko which have UV coverage. We treat luminosity estimates with an uncertainty larger than 1 dex as upper limits in this plot. The bars on the left y-axis correspond to the peak luminosity or the weighted average of radius or temperature for objects in that color. The same is shown in light blue for the non-repeating TDE comparison sample with the shading representing the 1$\sigma$ interval.}
 \label{fig:22ci_BB_fits}
\end{figure*}

We fit the time-evolving UV/optical SED of ASASSN-22ci with a blackbody model, using MCMC and forward modeling methods. We only fit epochs with \swift UVOT photometry to avoid issues in the cross-calibration between data sources and the need to interpolate data. These fits yielded estimates of the bolometric UV/optical luminosity, temperature, and effective radius for each epoch. A blackbody model describes the UV/optical emission of ASASSN-22ci well, with a mean (median) reduced $\chi^2$ of 4.6 (3.2), similar to other TDEs \citep[e.g.][]{holoien14b}.

As we have done for other TDEs \citep{holoien20, hinkle20a, hinkle21b}, we created a bolometric UV/optical light curve by scaling the ATLAS $o$ data to match the bolometric luminosity estimated from the fits to the \swift UVOT data. For epochs within the \swift coverage the bolometric correction is interpolated and for epochs outside of the \swift coverage we assume a constant bolometric correction, effectively assuming a fixed temperature. From this bolometric light curve, we computed the time of peak and the peak luminosity for each flare by fitting a smooth spline to the near-peak data of each flare and estimating the uncertainty from the standard deviation of 10,000 Monte Carlo iterations. For the first flare, we find a peak luminosity of $\textrm{log}_{10} (\textrm{L} \ [\textrm{erg } \textrm{s}^{-1}]) = 43.92 \pm 0.03$ on MJD $59635.1 \pm 0.9$. The second flare had a fainter peak of $\textrm{log}_{10} (\textrm{L} \ [\textrm{erg } \textrm{s}^{-1}]) = 43.57 \pm 0.05$ on MJD $60354.9 \pm 4.6$.

The results from the blackbody fits are shown in Figure \ref{fig:22ci_BB_fits}. We have excluded any fits to the data between the flares for ASASSN-22ci as these detections are typically $<3\sigma$ and the resulting blackbody parameters appear spurious. We show ASASSN-22ci along with a sample comparison objects composed of the other multiple/repeating TDE candidates ASASSN-14ko \citep{payne21,payne22,payne23}, ASASSN-18ul \citep[AT2018fyk;][]{wevers19, wevers23}, ASASSN-19dj \citep[AT2019azh;][see Section \ref{sec:multiples}]{hinkle21a}, and AT2020vdq \citep[ZTF20acaazkt;][]{somalwar23}. Compared to the background sample of TDEs \citep[][shown in light blue]{hinkle21b, hoogendam24}, the peak luminosity for the first flare of ASASSN-22ci is consistent with the average TDE peak luminosity when including faint and fast events. The peak luminosity of the second flare is lower than average, but well within the range of faint and fast TDEs. ASASSN-22ci has the lowest peak luminosity of any UV/optical TDE with multiple or repeated flares to date.

The blackbody radius of ASASSN-22ci is smaller than the average TDE effective radius, but only slightly smaller than the other multiple/repeated TDE candidates. The radius evolution is similar to other TDEs in that it decreases after peak as the emission fades. The blackbody temperature of ASASSN-22ci is hot at peak, with a temperature of $\sim$30,000 K for each of the flares which is hotter than the average TDE temperature during its evolution. ASASSN-22ci exhibits cooling during both of its flares with a temperature change that is much larger than typically observed for TDEs and opposite to the trend that some TDEs show increasing temperatures with time \citep{hinkle20a, vanvelzen21}. Strikingly similar behavior is seen for the flares of ASASSN-14ko, although on much shorter timescales. Such a trend does not appear to be common among the other multiple/repeating TDEs.

\begin{figure}
\centering
 \includegraphics[width=0.49\textwidth]{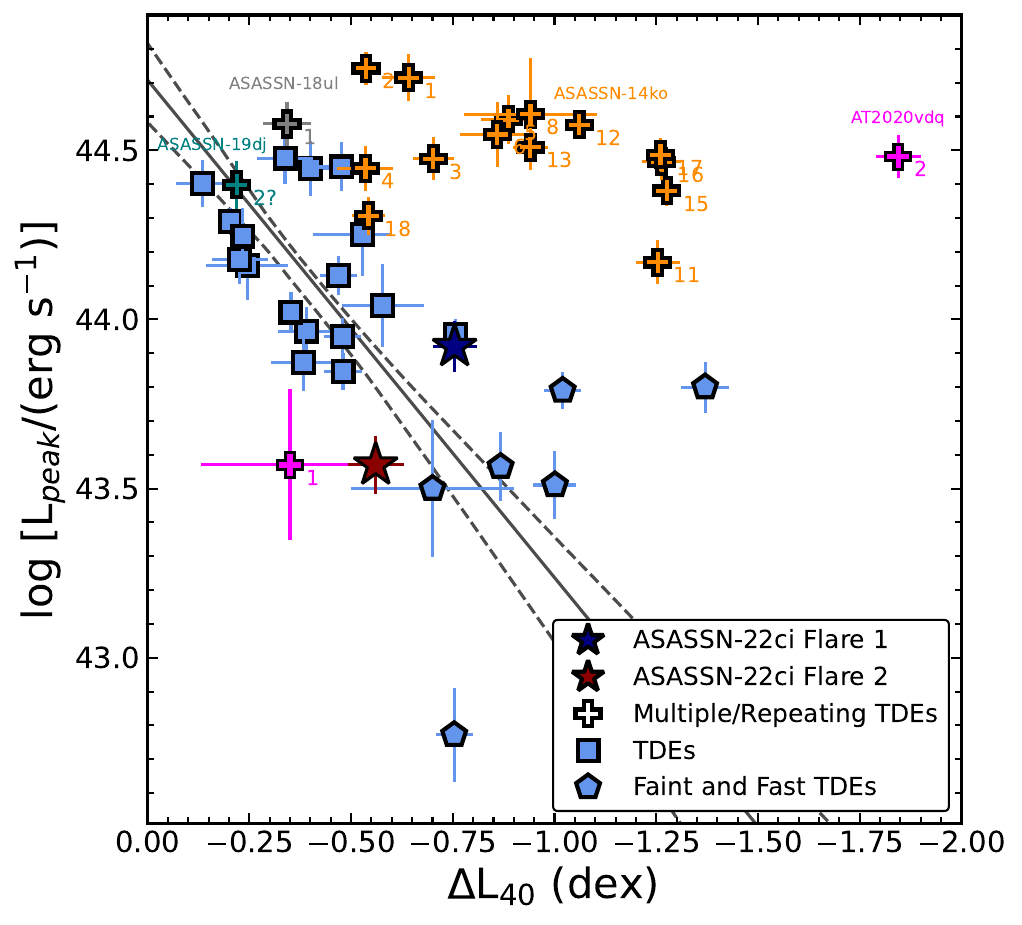}
 \caption{The peak luminosity of a TDE as compared to its decline rate. The decline rate $\Delta$L$_{40}$ is defined as the difference between the log of the peak luminosity and the log of the luminosity 40 days after peak. The two flares of ASASSN-22ci are shown as stars. The blue squares and pentagons correspond to ``normal'' TDEs and faint and fast TDEs, respectively \citep{hinkle20a, hinkle21b, hoogendam24}. The pluses are multiple/repeating TDEs using the same colors as in Fig. \ref{fig:22ci_BB_fits}. The flare number for each object is given at the bottom right of the point. The black solid line is the best-fit to the non-repeating TDEs with the allowed range of uncertainty given as the dashed black lines.}
 \label{fig:deltaL40}
\end{figure}

Additionally, we used the bolometric UV/optical light curves to compare the temporal evolution of ASASSN-22ci to other TDEs, including those with multiple or repeating flares. We follow the method of \citet{hinkle20a} using the decline in brightness at 40 rest-frame days post-peak. We also apply this method to AT2020vdq \citep{somalwar23} and the flares of ASASSN-14ko presented in \citet{payne21}. All of the TDEs with multiple/repeating flares are shown in Figure \ref{fig:deltaL40} as plus symbols using the same colors as in Fig. \ref{fig:22ci_BB_fits}.

The two flares for ASASSN-22ci lie reasonably close to the TDE peak-luminosity/decline rate relationship calculated in \citet{hinkle20a, hinkle21b}. The less luminous second flare of ASASSN-22ci atypically decays more slowly after peak than the more luminous first flare. A similar trend is seen for AT2020vdq, where the first flare is significantly less luminous and decays more slowly than the second flare. However, the first flare of AT2020vdq has no UV observations so the luminosity estimates assume a bolometric correction from the first \swift epoch during the second flare. We account for this unknown systematic by adding a 50\% uncertainty in quadrature with the estimated uncertainty on the peak luminosity of the first flare. The flares for ASASSN-14ko appear to follow a qualitatively similar trend to the ensemble of TDEs, with less luminous flares decaying more quickly, although the relationship is much shallower \citep{payne21}. As a population, many of the multiple/repeated TDE lie above the typical peak-luminosity/decline rate relationship, which may be expected given that these events are inferred to be partial TDEs, which are expected to decay more quickly \citep[e.g.,][]{coughlin19}.

\subsection{Mid-Infrared Constraints} \label{sec:MIR}

Finally, we obtained the NEOWISE \citep{mainzer11, NEOWISE_phot} $W1$- and $W2$- band light curves and searched for a mid-IR dust echo. We find no detected IR emission above that of the host galaxy at any point after the first flare begins. We estimate a conservative limit on the MIR emission as 3 times the sum of the most-permissive $W1$ and $W2$ flux limits. At the distance of ASASSN-22ci this yields a luminosity of $L_{MIR} < 2 \times 10^{41}$ erg s$^{-1}$. Using the simplistic approximation of the dust covering fraction as the ratio of the peak IR luminosity to the peak UV/optical luminosity, this corresponds to a limit on the covering fraction of $\lesssim 0.3$\%, consistent with many optically-selected TDEs \citep{vanvelzen16b, jiang21b} and in line with the lack of coronal lines in the optical and NIR spectra, assuming a typical gas-to-dust ratio in the circumnuclear medium. This limit is well below the typical covering fraction for ambiguous nuclear transients \citep[ANTs;][]{hinkle24b}, consistent with a lack of existing AGN activity in the host galaxy of ASASSN-22ci.

\subsection{X-ray Emission} \label{sec:xray}

As discussed in Section \ref{sec:xrt} and shown in Figure \ref{fig:xray_images}, we find no X-ray emission associated with ASASSN-22ci during either optical/UV flare. The lack of X-ray emission is common for optically-selected TDEs \citep[e.g.,][]{auchettl17, vanvelzen20b, hammerstein23, hoogendam24} and may suggest that the X-rays have been completely reprocessed by the surrounding stellar debris. Compared to the UV/optical bolometric luminosity detected at peak for both flares, the X-ray emission is 2-3 orders of magnitude lower, similar to other events where no X-ray emission is detected. Assuming the SMBH mass derived in Section \ref{sec:archival}, this suggests that during the flares, the X-ray emission is $\lesssim0.03-0.05\%$ of Eddington, which is consistent with what has been seen in other X-ray emitting TDEs \citep[e.g.,][]{auchettl17, wevers19, saxton20, wevers23, guolo24}.

However, what is unique about this source is that we detect significant X-ray emission between the flares. Late-time X-ray emission associated with optical/UV TDEs that showed no strong (or very weak) X-ray emission during the original flare is now commonly observed. For example, the optical TDE candidates PTF09axc, PTF09ge, ASASSN-14ae \citep{jonker20}, AT2019fdr \citep{reusch22}, OGLE16aaa \citep{shu20, kajava20}, AT2019qiz \citep{nicholl24}, AT2019vcb \citep{quintin23} and AT2023lli \citep{huang24} did not exhibit strong X-rays during the original flare but were detected as X-ray sources months to years after the original TDE flare had faded. This differs from events that exhibit significant X-rays during the optical/UV flare and then show a rebrightening at later times, like that seen in ASASSN-15oi \citep{gezari17, holoien18a, hajela24}, ASASSN-19dj \citep{hinkle20a, liu22}, or AT2018fyk \citep{wevers21, wevers19, wevers23b, pasham24}. 

\citet{hayasaki21} suggested that for the events that exhibited X-ray emission $>$2 years after the optical/UV flare, this was a natural consequence of circularization, the impact factor, and whether the accretion is sub-Eddington or not, with the time delay between the optical/UV and X-ray emission connected to the circularization and accretion timescales. For events like AT2023lli, which exhibited significant X-ray emission immediately after the optical/UV declined, \citet{huang24} and \citet{wevers23b} suggested this could have resulted from the obscuring material or accretion disk becoming radiatively inefficient and/or optically thin, and any evidence of weak X-ray emission during the initial flare could be due to inhomogeneously distributed obscuring material. More recently, AT2019qiz and AT2019vcb also exhibited late-time X-ray emission in the form of Quasi-periodic Eruptions \citep[QPEs;][]{quintin23, nicholl24}, which are bright bursts of X-ray emission that repeat on timescales of hours to weeks. It is thought that these can either arise from instabilities associated with the accretion disk or due to the interaction of a tightly bound stellar object with the accretion disk. 

To place constraints on the nature of the X-ray emission seen between the flares of ASASSN-22ci, we first calculated a hardness ratio, $HR$ = (H - S)/(H + S), to determine the ratio of hard (H; 2.0-10.0 keV) to soft (S; 0.3-2.0 keV) X-ray emission. We find a hardness ratio of $-0.49\pm0.35$, suggesting that the emission is relatively soft, similar to the predominately thermal X-ray emission of other TDEs \citep[e.g.,][]{brown17a, auchettl17, holoien18a, auchettl18, wevers21, wevers19, wevers23, guolo24}. Fitting the low count spectrum with an absorbed blackbody redshifted to the host galaxy, we find that the emission is well fit by a temperature of $0.042\pm0.01$ keV and a blackbody radius of $(4.7^{+6.8}_{-2.8})\times10^{11}$ cm. We find that fitting the column density as well does not improve the fit and so we fix it to the Galactic column density ($N_{H}=1.92\times10^{20}$ cm$^{-2}$; \citealt{HI4PI16}).

The derived temperature and radius are similar to that found for other TDEs such as ASASSN-14li \citep[e.g.,][]{brown17a}, ASASSN-15oi \citep[e.g.,][]{gezari17, holoien18a, hajela24}, ASASSN-19dj \citep[][]{hinkle21a} and some of those found in \citet[][see Figure 5 or Table 7]{guolo24}. Assuming a disk that exhibits no absorption or/and reprocessing, we find that the radius of the blackbody component is consistent with the radius of the innermost stable circular orbit, similar to the findings of \citet{mummery23} and \citet{guolo24}. The total unabsorbed 0.3-10.0 keV X-ray luminosity is $(3.5\pm1.3)\times10^{41}$ erg s$^{-1}$, which corresponds to an Eddington ratio of $\sim$0.001. This is lower than seen in X-ray bright TDEs \citep{auchettl17, mockler19, guolo24} and lower than that derived for AT2019qiz whose X-ray emission was originally veiled, but exhibited quasi-periodic X-ray emission at late times, well after the original optical flare decayed \citep{nicholl24}. 

Assuming that the properties of X-ray emission seen between the two flares are consistent with the X-ray emission associated with the two flares, we can estimate the column density needed to veil this X-ray emission. The necessary column densities are $N_{H} \sim 0.9\times10^{21}$ cm$^{-2}$ and $N_{H} \sim 1.3\times10^{21}$ cm$^{-2}$ for Flare 1 and 2. This is $\sim$$4 \mbox{ -- } 6$ times greater than the Galactic column density along the line of sight, consistent with \citet{auchettl17}, who found that nearly all TDEs are highly absorbed.

\section{Insights from Events With Multiple Flares} \label{sec:multiple_flares}

\begin{figure*}
\centering
 \includegraphics[height=0.47\textwidth]{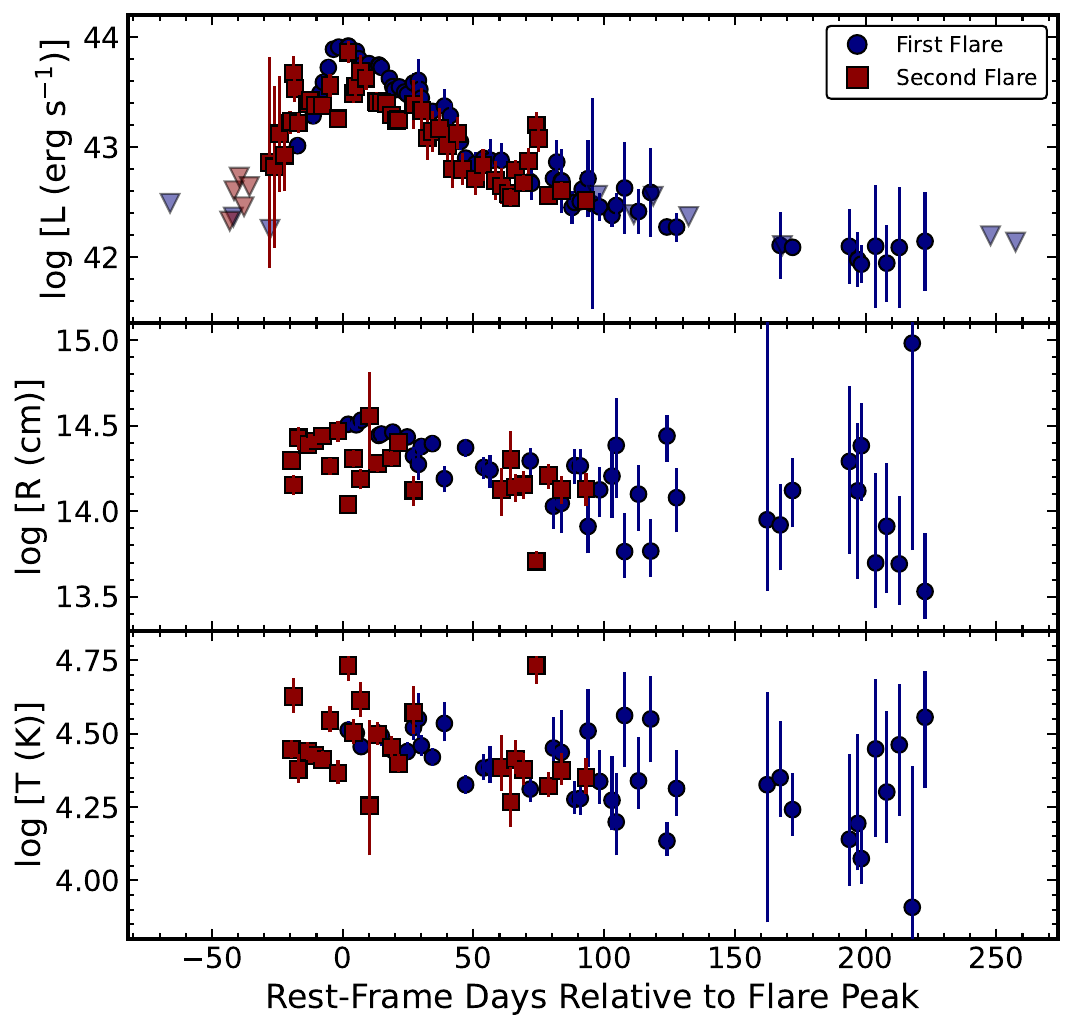}
  \includegraphics[height=0.47\textwidth]{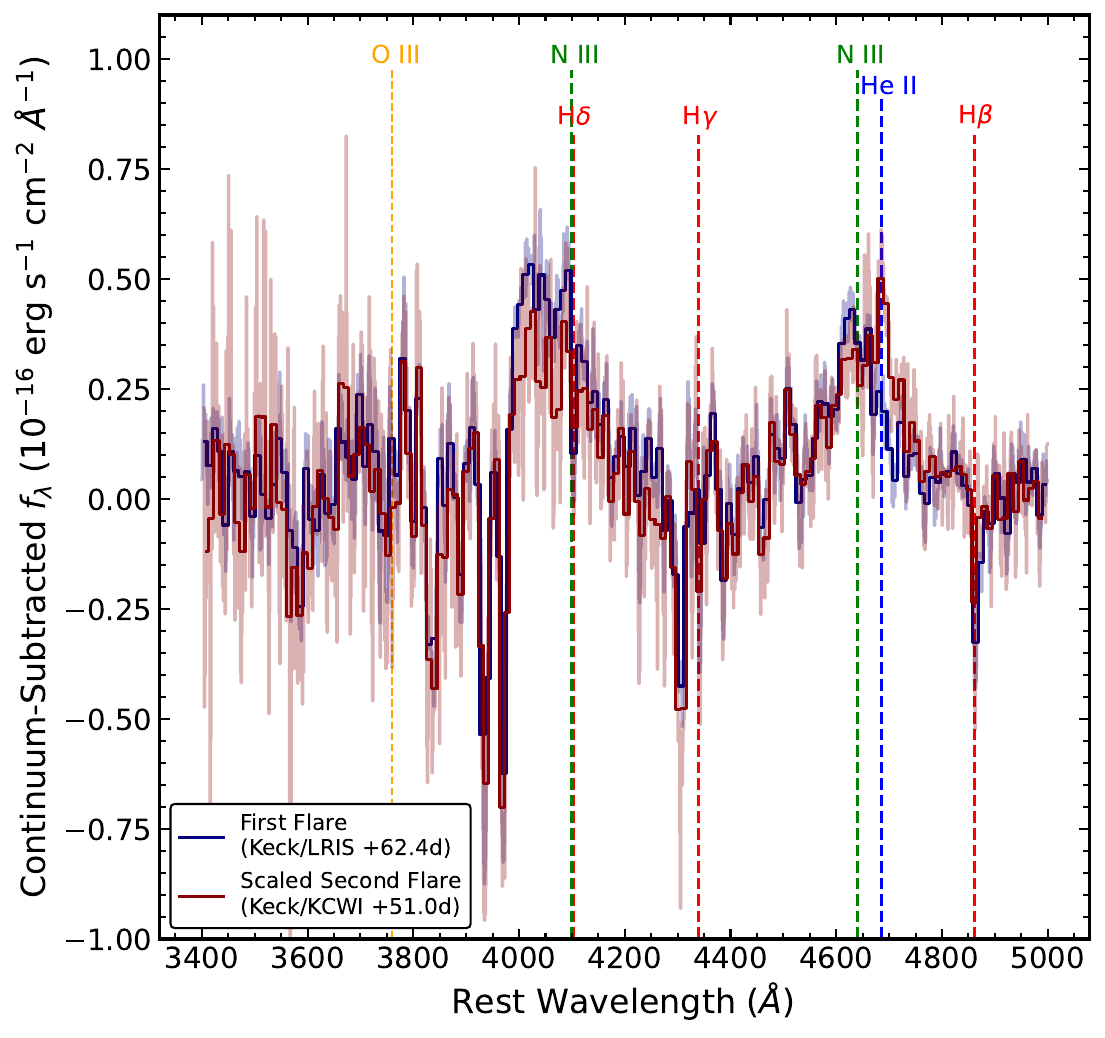}
 \caption{\textit{Left Panel:} Comparison of the bolometric UV/optical luminosity (top), effective radius (middle), and effective temperature (bottom) of ASASSN-22ci from blackbody fits to the UV/optical SEDs of the flares. The first flare (blue) and second flare (red) are each shown relative to their peak time in rest-frame days. We treat luminosity estimates with an uncertainty larger than 1 dex as upper-limits, denoted with downward-facing triangles. \textit{Right Panel:} Comparison of the continuum-subtracted spectra of the two flares. Shown are blue portions of the spectra taken with Keck at similar phases from the flare peaks. The KCWI spectrum has been scaled by the $\approx$1.1 mag difference in the synthetic $B$-band transient photometry between the two epochs, as the second flare is fainter than the first. Common TDE emission features are marked, using the same colors as in Fig. \ref{fig:opt_spec}.}
 \label{fig:flares_comp}
\end{figure*}

\subsection{Comparison of the Two Flares from ASASSN-22ci}

ASASSN-22ci is one of only five TDEs that have been claimed to have multiple flares. It is even rarer in that both of its flares have good coverage with multi-wavelength photometry and high S/N optical spectroscopy. This allows us to compare the properties of the two flares in more detail than is possible for nearly any event other than ASASSN-14ko \citep[e.g.,][]{payne23}. Given the quiescent Balmer strong nature of the host galaxy of ASASSN-22ci, this galaxy may have a TDE rate that is $20 \mbox{--} 30\times$ the TDE rate in an average galaxy \citep[][see Fig.\ 7 of \citealt{french20b}]{french16, graur18}. While unlikely, we cannot rule out that we are simply viewing two distinct TDEs, although such a high rate of TDEs would be interesting in its own right.

However, a comparison of the flare properties gives us additional insight into this possibility. The left panel of Figure \ref{fig:flares_comp} compares the blackbody parameters for the two flares of ASASSN-22ci. It is immediately clear that the two flares are remarkably similar in nearly all regards. For both flares, the early-time emission is detectable beginning $\sim$$25$ to $20$ days prior to peak. The overall flare shapes are quite similar, although the second flare rises slightly more slowly and peaks at roughly half the luminosity of the first flare. The peak is also slightly flatter for the second flare, consistent with its position in Fig.\ \ref{fig:deltaL40}. Nevertheless, these are minor differences --- the two flares of ASASSN-22ci are much more alike than any random pairing of two TDE flares. The radius and temperature evolution of the two flares are nearly identical, with similar values at peak and temporal trends. Both flares have hot and compact photospheres throughout their observed evolution, suggesting a similar structure in their accretion flows. A similar SED behavior across multiple flares is also seen for ASASSN-14ko \citep[][also see Fig. \ref{fig:22ci_BB_fits}]{payne23}.

Given that both flares arose from the same SMBH, a similar flare evolution might be expected regardless of whether the flares of ASASSN-22ci result from a repeating disruption or two separate flares. However, TDE population studies have found no strong trends between the stellar mass of the host galaxy (a proxy for SMBH mass) and the blackbody temperature or radius at peak \citep{vanvelzen21, hammerstein23}. Additionally, for TDEs with a similar host-galaxy mass to ASASSN-22ci, there is nearly 0.5 dex spread in both the blackbody radii and temperatures. Furthermore, even though the SMBH mass and spin are unchanged between the two flares, the mass, evolutionary stage, and impact parameter of the star and its orbit relative to the SMBH spin will differ, leading to different flare properties \citep[e.g.][]{gafton19, mockler19}. Given the similarities between the flares, it is likely that ASASSN-22ci results from a single star being partially disrupted twice.

The optical spectra of the two flares are also strikingly similar after correcting for the flux difference between the two flares. The right panel of Figure \ref{fig:flares_comp} shows two spectra taken at a rest-frame phase of $\sim$$+50 \mbox{ -- } 60$ days from peak. After subtracting the local continuum, the two spectra show the same line features. Each flare exhibits weak \ion{O}{3} emission with much stronger emission from \ion{N}{3} and the \ion{N}{3}/\ion{He}{2} blend. After the continuum flux scaling, the line shapes and strengths are also extremely similar.

The spectra of optically-selected TDEs have a wide diversity \citep[e.g.,][]{leloudas19, vanvelzen21, hammerstein23}, and while some are similar, no two are identical. Therefore, the nearly identical spectra seen for the two flares of ASASSN-22ci indicate an extremely similar gas environment immediately surrounding the SMBH. This suggests that the gas stripped to power each flare is similar, suggesting the repeated disruption of a single star. This is only the second TDE, after ASASSN-14ko, for which sufficient spectroscopic data exists over multiple flares to support such a claim. In addition to the optical spectra, the blackbody properties of the two flares of ASASSN-22ci are remarkably similar.

\begin{figure*}
\centering
 \includegraphics[width=0.99\textwidth]{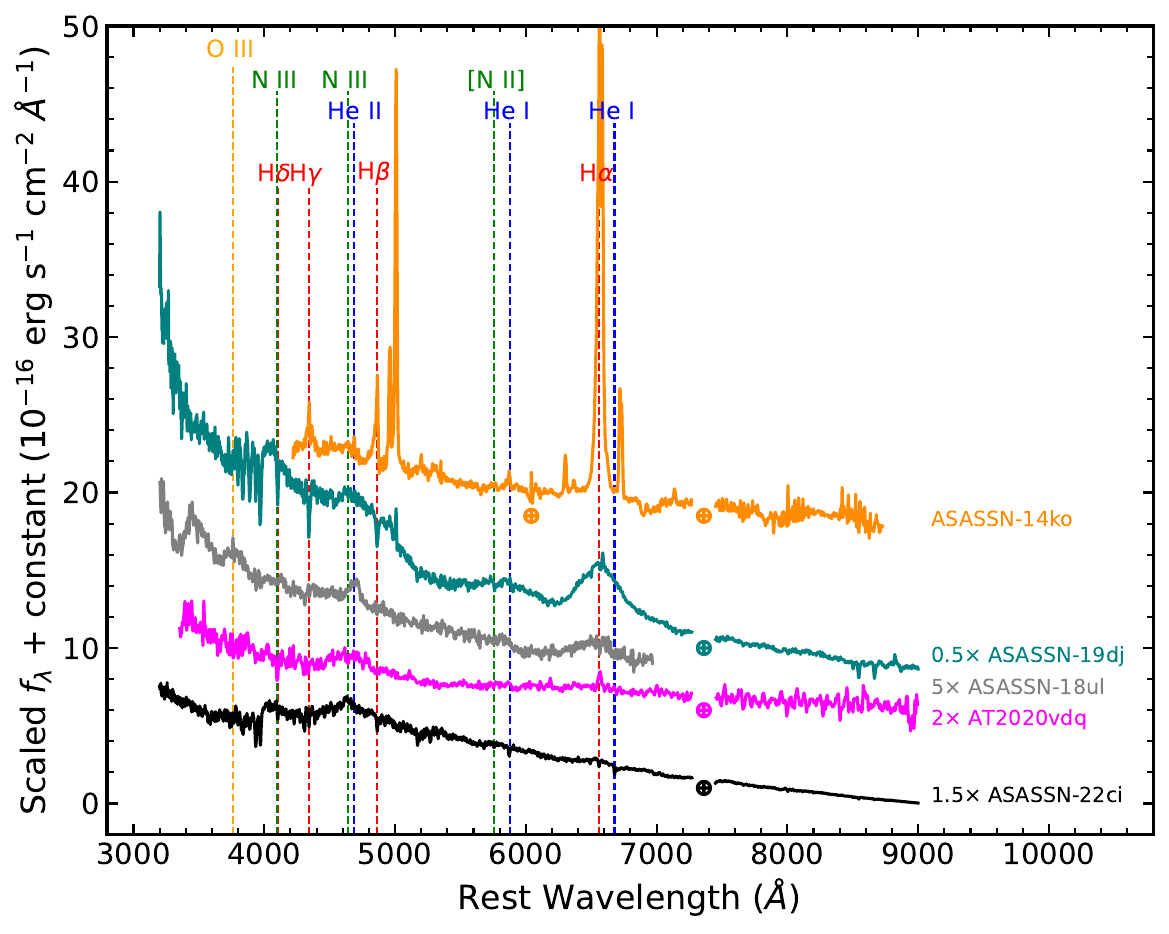}
 \caption{Comparison of the optical spectrum of ASASSN-22ci (black) to the multiple/repeating TDE candidates ASASSN-14ko (orange), ASASSN-18ul (gray), ASASSN-19dj (teal) and AT2020vdq (magenta). Each spectrum is scaled by the amount shown and plotted with an arbitrary offset to aid in visualization. The spectra are at a phase of $\sim$$+60$d relative to peak except for the faster-evolving ASASSN-14ko, which is at $\sim$$+15$d. Common TDE emission lines are marked with vertical dashed lines. Atmospheric telluric features are marked with an $\oplus$. }
 \label{fig:comparison_spec}
\end{figure*}

It is known that AGNs can exhibit coherent TDE-like flares \citep[e.g.,][]{graham17, auchettl18}. Although there are strong constraints on the presence of AGN activity for the host galaxy of ASASSN-22ci, we examined the possibility that the two flares seen for ASASSN-22ci might result from DRW-like variability \citep{giveon99, collier01, kelly09, zu11, zu13}. We used Javelin \citep{zu11, zu13} to generate 10,000 mock AGN light curves from a DRW model, with $\sigma$ and $\tau$ spanning the range of observed DRW parameters for AGNs \citep[e.g.,][]{kelly09, macleod12} and estimated the probability that flares similar to ASASSN-22ci could be produced. We first compared the RMS variability amplitude and skew of the mock light curves to the ATLAS $o$-band light curve of ASASSN-22ci and found that only 0.8\% have an RMS variability as high as ASASSN-22ci and $<0.1\%$ are as highly skewed. We additionally searched for light curves with a flare morphology similar to ASASSN-22ci, specifically selecting based on the peak flare amplitude relative to the baseline variability and flare width relative to the observed baseline. We counted mock DRW light curves which showed two or more flares with a fractional peak amplitude at least 75\% that of the fainter flare of ASASSN-22ci and two or more flares with fractional widths within 25\% of that observed for ASASSN-22ci. None of the 10,000 mock DRW light curves produced a light curve with two high-amplitude flares similar to ASASSN-22ci. Combined with the other evidence against an AGN in the host galaxy, it is extremely unlikely that these flares are due to AGN variability.

\subsection{Other TDEs with Multiple Flares} \label{sec:multiples}

ASASSN-22ci is not the first TDE to have shown multiple luminous flares. To date, the other optically-selected candidate\footnote{\citet{somalwar23} also find a second flare for the TDE AT2021mhg, but follow-up photometry and spectra indicate that the second flare is better explained by a Type Ia supernova.} examples are ASASSN-14ko \citep{payne21, payne22, payne23}, ASASSN-18ul \citep{wevers19, wevers23}, AT2020vdq \citep{somalwar23}, and ASASSN-19dj \citep{hinkle21a}. The TDE ASASSN-19dj had a nuclear flare in CRTS data roughly 14.5 years prior to the TDE detected by ASAS-SN \citep{hinkle21a}. Given the lack of a spectrum or multi-wavelength constraints, there is no definitive classification. Nevertheless, the discovery of several multiple/repeating TDEs with a range of recurrence periods increases the likelihood that this was a previous TDE, possibly of the same star. Additionally, AT2019aalc \citep{veres24}, AT2021aeuk \citep{sun25}, and flares in IRAS F01004-2237 \citep{sun24} have been claimed to be repeating partial TDEs. However, given that their host galaxies also host an AGN and thus have a higher probability of false positives due to AGN flares, we will not consider them for the remainder of the manuscript. There are also several repeating TDE candidates discovered in the X-ray, eRASSt J045650.3–203750 \citep{liu23, liu24_eRASSt} and RX J133157.6-324319.7 \citep{malyali23}. RX J133157.6-324319.7 shows no UV/optical emission and eRASSt J045650.3–203750 is only weakly variable in the UV with no optical variability. Of course, other TDEs that are currently considered single events may exhibit a second flare in the future. Such events will be important to constrain the period distribution of repeating TDEs.

Fig.\ \ref{fig:22ci_BB_fits} shows the blackbody fits for these optically-selected multiple/repeating TDE candidates. There is a wide range in the peak luminosities of these TDEs, with ASASSN-22ci being among the faintest. Nevertheless, each of the multiple/repeating TDEs is more luminous at peak than the average TDE in our comparison sample. A Kolmogorov–Smirnov (KS) test yields a probability of 0.1\% of the null hypothesis that the multiple/repeating TDE peak luminosities are drawn from the same distribution as our comparison sample. This is intriguing given that partial TDEs are generally expected to be less luminous than full TDEs for similar stellar and SMBH parameters \citep[e.g.,][]{law-smith20, liu24_simulation}. 

As shown in Fig.\ \ref{fig:deltaL40}, different flares of the multiple/repeating TDEs do not appear to follow the general peak-luminosity/decline-rate relationship of TDEs. In fact, some events like ASASSN-22ci and AT2020vdq show the opposite behavior, with their less luminous flares decaying more slowly after peak. Despite being offset from the locus of normal TDEs in this parameter space, the flares of ASASSN-14ko do show a weak trend where the more luminous flares decay more slowly. Consistent with their presumably partial TDE nature \citep[e.g.,][]{coughlin19, bandopadhyay24}, the decline rates of most of the multiple/repeating TDE flares are faster than is typical for their peak luminosity. 

Interestingly, the effective temperatures of the multiple/repeating TDEs generally seem to be hotter than an average TDE and their effective radii are slightly smaller (see Figure \ref{fig:22ci_BB_fits}). While suggestive, both parameters are nevertheless consistent with the $1\sigma$ range for typical optically-selected TDEs and a KS test cannot distinguish between the two populations. ASASSN-18ul is unique among these multiple/repeating TDE candidates in that it shows significant short-term variability in its light curve, atypical of TDEs \citep[e.g.,][]{hinkle20a, vanvelzen21}.

A comparison of the optical spectra of the multiple/repeated TDEs is shown in Figure \ref{fig:comparison_spec}. The obvious outlier here is ASASSN-14ko, whose spectra are significantly contaminated by common AGN emission features from the host-galaxy AGNs \citep{tucker21}. All objects show H$\alpha$ emission, with the weakest line seen for ASASSN-22ci. Notably, all spectra also show the broad \ion{He}{2} $\lambda4686$ and/or \ion{N}{3} $\lambda4640$ feature. ASASSN-22ci, ASASSN-19dj, and ASASSN-18ul show clear \ion{N}{3} $\lambda4100$ emission. AT2020vdq shows no strong \ion{N}{3} $\lambda4100$ line but does exhibit N emission in the UV \citep{somalwar23}. This means that all of the optically-selected multiple/repeating TDEs are of the TDE-H+He class and likely TDE-H+He (N). Given the 17 TDE-H+He out of a total of 30 TDEs in \citet{hammerstein23}, the binomial probability of 4 out of 4 multiple/repeating TDEs being TDE-H+He is 10\%. At 1.3$\sigma$ this is currently explainable by random chance, but trends in the spectra of newly-discovered multiple TDEs should be monitored closely.

Like normal optically-selected TDEs, the multiple/repeating TDEs are frequently hosted by galaxies with strong Balmer absorption, but weak emission lines. This is often a signature of a post-starburst system where there remains significant light from A stars but without the ionizing flux of OB stars \citep[e.g.,][]{french16, french18}. These are also the galaxies for which the TDE rates are the highest, with some systems likely to have TDE rate enhancements of $\sim$100 times the average galaxy \citep[e.g.][]{french16, law-smith17, graur18}. With these galaxies potentially hosting TDE rates of $\gtrsim$$7\times10^{-3}$ yr$^{-1}$ \citep[e.g.,][]{yao24}, it is significantly more likely that some subset of the multiple/repeating TDEs are unrelated events and not multiple disruptions of the same star. Continued follow-up of these events is required to search for future flares and place constraints on the true fraction of TDEs with repeating flares. Unlike the other multiple/repeating TDEs, ASASSN-14ko resides within a galaxy that is a merger remnant and hosts a strong AGN \citep{tucker21}.

\section{Discussion and Conclusions} \label{sec:disc_conc}

The sample of TDEs with multiple flares is growing rapidly. We have now observed 5 optically-selected TDEs that exhibit multiple flares, some of which are likely to be repeating partial tidal disruptions. Although small, this sample is nevertheless sufficient to begin exploring the theoretical implications of the observed trends among these events. 

First, we examine the observed properties of each of the proposed UV/optical multiple/repeating TDEs. ASASSN-14ko has shown clear repeating flares for the entirety of ASAS-SN survey operations \citep{payne21, payne22, payne23} and has characteristics consistent with a repeating partial TDE \citep[e.g.][]{payne21, linial24, bandopadhyay24}. ASASSN-18ul shows signs of periodic activity \citep{wevers23, pasham24}, although the significant optical variability during the flare, relatively hard X-ray emission, and high SMBH mass are atypical of TDEs. While AT2020vdq does exhibit two flares, the first flare was not observed in the UV or X-ray and there is no spectrum available. Thus, while it is plausibly consistent with a TDE, we cannot make a definitive statement. Furthermore, the light curve shapes of the two flares from AT2020vdq are distinct, particularly with respect to their decay timescale. Similarly, the earlier flare observed for ASASSN-19dj has no spectrum or multi-wavelength information and cannot be conclusively classified as a TDE. Finally, ASASSN-22ci has two well-observed flares, with each event having strong evidence for being powered by a TDE. We will consider each of these candidate repeating TDEs for the sake of this discussion, but it would not be surprising if some are not true examples.

\begin{figure*}
\centering
 \includegraphics[width=0.99\textwidth]{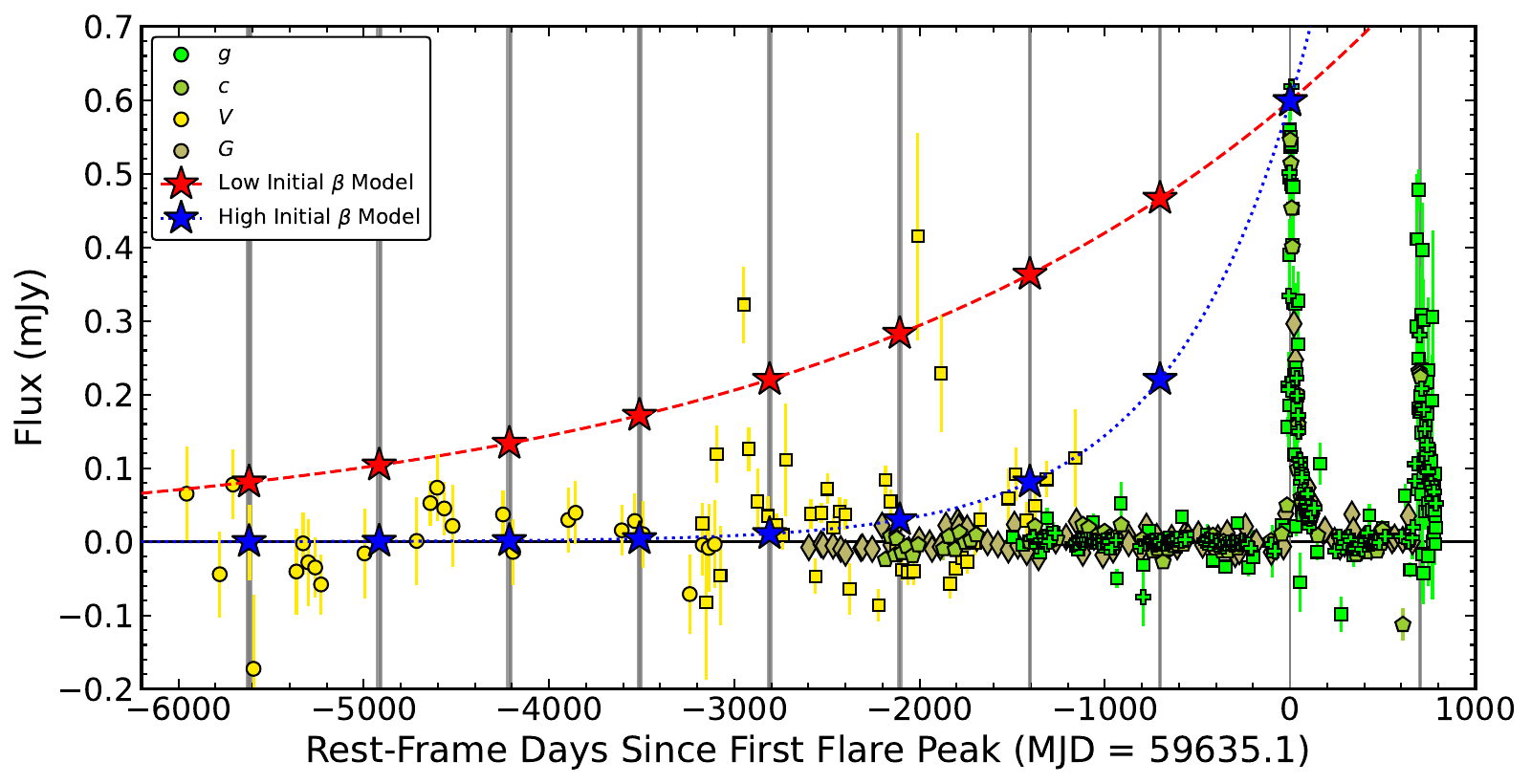}
 \caption{Long-term host-subtracted and foreground extinction-corrected light curve for ASASSN-22ci, with data from CRTS ($V$, yellow circles), ASAS-SN ($gV$, green and yellow squares), Gaia ($G$, khaki diamonds), ATLAS ($c$, yellow-green pentagons), and ZTF ($g$, green pluses). The horizontal black solid line represents zero flux and the vertical gray bands represent the observed flare recurrence time projected into the past, with the width of the band representing the uncertainty on the time of the expected flare peak.  Two repeating TDE flare models, with the expected flare times and exponentially rising peak fluxes are shown as stars. The red case represents a low initial $\beta$, with a slower increase in peak flux per flare and the blue case represents a high initial $\beta$ with a steeper rise. The pre-flare data strongly rules out any prior flares at a level consistent with either model prediction.}
 \label{fig:repeating_model}
\end{figure*}

\subsection{Constraints on Previous Flares}

We systematically searched for evidence of flares prior to the initial discovery. ASASSN-14ko, the source with the strongest likelihood of being a repeated TDE, has consistently shown flares throughout the ASAS-SN coverage \citep{payne21} and its first flare was likely not observed. For ASASSN-18ul, we find no evidence of a previous flare in ASAS-SN data and so it is likely that we detected the first flare. AT2020vdq has no flares in earlier ASAS-SN or CRTS \citep{somalwar23} data and is also likely the first flare. The long required period for the flares of ASASSN-19dj makes it impossible to determine whether the flare seen in CRTS data was the first. Finally, Fig.\ \ref{fig:long_lc} shows that we detected the first flare of ASASSN-22ci. In total, of the 4 proposed repeating TDEs for which a constraint can be made, 3 have likely been observed on their first flare.

This is somewhat surprising, as most models of repeating TDEs predict multiple flares before the star is ultimately fully disrupted \citep[e.g.,][]{liu24_simulation, bandopadhyay24}. However, the evolution of repeating TDE flares should vary strongly with the evolutionary state of the disrupted star because a star becomes more centrally concentrated and difficult to fully disrupt as it evolves \citep[e.g.,][]{bandopadhyay24}. In the case of a main sequence star, the star is expected to experience progressively deeper and deeper disruptions powering more and more luminous flares over time. A critical parameter in this scenario is the initial impact parameter ($\beta$) of the first TDE. For a weak initial encounter, little mass is stripped from the star and the star can survive $\sim$10 orbits before the tidal heating and mass loss are significant \citep{liu24_simulation}. However, for an initial impact parameter close to 1, the star may only survive a small handful of disruptions before being fully disrupted. This model qualitatively describes the more luminous second flares for AT2020vdq and ASASSN-19dj, but not the less luminous second flares for ASASSN-18ul and ASASSN-22ci. Conversely, an evolved star is able to lose only a small amount of mass over many tens of orbits and should produce flares of similar amplitude with regular spacing, as seen for ASASSN-14ko \citep[e.g., see][]{liu24_simulation}.

For the simpler case of the evolved star repeating TDEs, we calculate the likelihood that we would have detected the first flares for 3 of the 4 repeating TDEs with the necessary data to place such constraints. Assuming that each star experiences 10 (20) successive partial TDEs, each with a similar peak brightness, the binomial probability of finding 3 of 4 on their first flare is extremely low at 0.4\% (0.05\%). For $\sim$100 repeated flares, the probability of catching this many first flares is vanishingly small. In such a scenario, as long as the flares peak at a similar luminosity, any earlier flares should easily be detectable provided the period is short enough relative to the survey baseline.

The main sequence case is more complicated, as the peak brightness of the flares is expected to evolve strongly with each flare. Furthermore, the simulations of \citet[][c.f.\ their Fig.\ 2]{liu24_simulation} found that the mass-loss rate increases with the number of pericenter passages more slowly ($\dot{M} \propto e^{N/4}$) for low initial impact parameters ($\beta = 0.5$) and more rapidly ($\dot{M} \propto e^{N}$) for initial impact parameters close to unity, with a transition near $\beta = 0.6$. Using these scalings, we conducted a simple simulation to test these predictions with respect to the detection of repeating TDEs. We generated a mock TDE flare profile by smoothing the ATLAS $o$-band data and generated a light curve of a mock repeating TDE with a given number of flares, time of separation, and luminosity evolution per passage.

We generated a synthetic long-term evolution of ASASSN-22ci, scaling the final flare to be the observed peak flux of the first flare in the appropriate band. For the low initial $\beta$ case, with a $e^{N/4}$ peak evolution, ASAS-SN would have been sensitive to 5 previous flares. For the steeper $e^{N}$ peak evolution, more appropriate for an initial $\beta$ near 1, ASAS-SN would have detected only 1 flare prior to the initial detection. Although discovered by ASAS-SN, ASASSN-22ci also has ATLAS and ZTF data that constrain the presence of flares before 2022. For the low (high) initial $\beta$ case we find that ATLAS would have been sensitive to 7 (1) prior flares and that ZTF would have been sensitive to 12 (3) previous flares. Figure \ref{fig:repeating_model} shows the predictions of these models, with the low initial $\beta$ in red and high initial $\beta$ in blue. For either model, the brightest previous flare should have occurred one orbit before the initial detection of ASASSN-22ci. The survey light curves strongly rule out any flare at this time (or any previous expected flare time), regardless of which model we assume.

These models suggest a potential problem in the calculated rates of TDEs. Repeating TDEs have multiple chances of being detected. Therefore, without the knowledge that a TDE is repeating, which is especially problematic for long flare recurrence times, the TDE rate can be overestimated by a factor up to the average number of repeated flares per TDE. Continued follow-up of known TDEs will allow us to examine the full range of flare recurrence timescales and repeating TDE behaviors.

\begin{figure*}
\centering
 \includegraphics[width=0.9\textwidth]{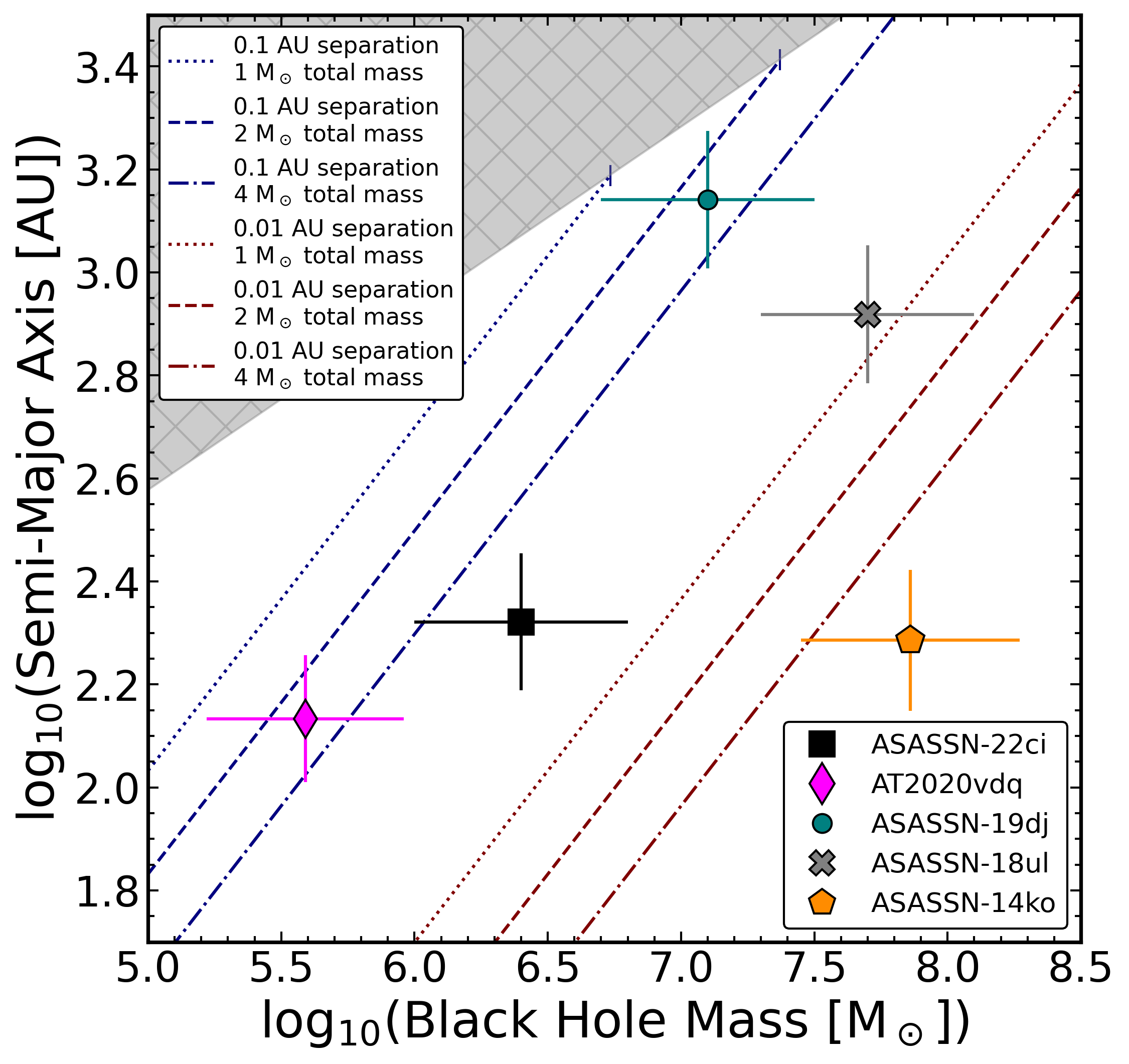}
 \caption{Semi-major axis of the bound star as compared to the central SMBH mass for our sample of multiple/repeating TDE candidates, assuming Hills capture. Shown are ASASSN-22ci (black square), AT2020vdq (magenta diamond), ASASSN-19dj (teal circle), ASASSN-18ul (gray cross), and ASASSN-14ko (orange pentagon). The lines correspond to initial binaries with different total masses and separations. The blue (red) lines are binaries with separations of 0.1 AU (0.01 AU). The dotted/dashed/dash-dotted lines are binaries with a total mass of 1/2/4 M$_{\odot}$, respectively. The gray hatched region in the upper left denotes a region that cannot be populated since the initial binary would not survive the velocity dispersion within the galactic bulge, assuming a separation of 0.1 AU and a typical M-$\sigma$ relation. The blue dotted and dashed lines are thus terminated at this limit.}
 \label{fig:rpTDE_space}
\end{figure*}

\subsection{The Parameter Space of Observed Repeating TDE Candidates}

We can also use the properties of the multiple/repeating TDE candidates to understand the population of binaries from which they arise. To do this, we first assume that the rest-frame flare recurrence time is the orbital period of the bound star. Using the orbital period and SMBH mass estimates, we calculated the corresponding semi-major axis of the star on its orbit around the SMBH. Figure \ref{fig:rpTDE_space} compares the implied semi-major axes and SMBH masses of the multiple/repeating TDE candidates to several initial binary configurations. For each combination of initial binary separation and mass, we calculate the semi-major axis of the bound star following Hills capture \citep[e.g.,][]{antonini11, cufari22}. In some cases, the initial binary would not be hard enough to survive within the galactic bulge, assuming a typical M-$\sigma$ relation \citep[e.g.,][]{gultekin09}.

As previously noted, the eccentricities of the bound stars powering these repeating flares must be extremely high \citep{cufari22, bandopadhyay24}. Here, we calculate the eccentricities of the stars powering the multiple/repeating TDE candidates by taking the orbital pericenter to be the partial disruption radius, assumed here to be twice the tidal radius (a $\beta = 0.5$ disruption) and comparing it to the semi-major axis. For a Solar mass main sequence star, the eccentricities range from 0.980 for ASASSN-14ko to 0.998 for ASASSN-19dj. All sources but ASASSN-14ko have implied eccentricities above 0.99, fully consistent with theoretical expectations \citep[][]{cufari22}. All main sequence stars require highly eccentric orbits, but significantly evolved red giant branch or asymptotic giant branch stars can power repeating TDE flares on moderately eccentric orbits. However, for the binary separations consistent with the multiple/repeating TDE candidates, highly evolved stars will have undergone common envelope evolution \citep{ivanova13}, likely making the binary harder to disrupt. Nevertheless, events like ASASSN-14ko appear to require evolved stars to explain their observed properties.

From Figure \ref{fig:rpTDE_space}, we find that all of the multiple/repeating TDE candidates are consistent with coming from binaries with a total mass between $\sim$$1 \mbox{ -- } 4$ M$_{\odot}$ and separations of $\sim$$0.01 \mbox{ -- } 0.1$ AU ($\sim$$2 \mbox{ -- } 20$ R$_{\odot}$). ASASSN-14ko lies slightly below the lines for our initial binaries, consistent with the estimated binary separation of 0.005 AU from \citet{cufari22}. We note that there is a degeneracy between the initial binary mass and separation in terms of the semi-major axis of the bound star. Nevertheless, the position of these sources is consistent with the disrupted stars being intermediate mass, potentially consistent with the presence of nitrogen in each of the multiple/repeating TDE spectra \citep{kochanek16a, mockler22}. While a partial TDE should only strip envelope material, processes like dredge-up in evolved stars \citep[e.g.,][]{vandehoek97} or rotational mixing \citep[e.g.,][]{deMink09, deMink13} can provide enhancements of CNO-processed material near the stellar surface.

\subsection{Future Flare(s) of ASASSN-22ci}

To date, ASASSN-22ci has exhibited two flares, with an observed separation between peaks of $720 \pm 4.7$ days. While it is unknown if this event is truly repeating, we will have the opportunity to test such a model in the coming years. If we assume that the rest-frame peak separation of 700 days is the orbital period of a star that survived the initial tidal encounter, then we would expect another flare to occur near MJD 61075 (2026 February 04). Follow-up observations starting before this date will confirm if this event is repeating on a timescale consistent with the observed flare separation and will provide a detailed look at the rising stages of a TDE with otherwise typical SED and spectral properties.

Repeating TDEs like ASASSN-22ci provide a unique opportunity to observe the earliest phases of TDE emission in detail. With additional flares from the same source, we can improve timing models and refine our ability to observe their rising light curves. With upcoming surveys like the Legacy Survey of Space and Time \citep{ivezic08}, we can directly test the predictions of models like those shown in Fig.\ \ref{fig:repeating_model} and understand the early-time evolution of repeating TDEs resulting from weak tidal encounters. As the sample of multiple/repeating TDE candidates continues to grow, so do our opportunities to leverage these objects to probe the populations of hard binaries in galactic centers, test TDE models, and constrain the UV/optical emission mechanism at these early phases.

\begin{deluxetable*}{ccccc}
\tablewidth{240pt}
\tabletypesize{\footnotesize}
\tablecaption{Spectroscopic Observations of ASASSN-22ci}
\tablehead{
\colhead{MJD} &
\colhead{UTC Date} &
\colhead{Telescope} &
\colhead{Instrument} & 
\colhead{Wavelength Range (\AA)}}
\startdata
\hline \multicolumn{5}{c}{First Flare}\\ \hline
59631.6 & 2024-02-21 & Faulkes Telescope North & FLOYDS & $3500 \mbox{ -- } 10000$ \\
59634.6 & 2024-02-24 & Keck II & KCWI & $3500 \mbox{ -- } 5700$ \\
59638.5 & 2022-02-28 & University of Hawai`i 2.2-m & SNIFS & $3400 \mbox{ -- } 9100$ \\
59651.5 & 2022-03-13 & University of Hawai`i 2.2-m & SNIFS & $3400 \mbox{ -- } 9100$ \\
59653.5 & 2022-03-15 & University of Hawai`i 2.2-m & SNIFS & $3400 \mbox{ -- } 9100$ \\
59655.4 & 2022-03-17 & University of Hawai`i 2.2-m & SNIFS & $3400 \mbox{ -- } 9100$ \\
59657.4 & 2022-03-19 & University of Hawai`i 2.2-m & SNIFS & $3400 \mbox{ -- } 9100$ \\
59661.0 & 2022-03-23 & Large Binocular Telescope & MODS & $3750 \mbox{ -- } 9000$ \\
59664.5 & 2022-03-26 & University of Hawai`i 2.2-m & SNIFS & $3400 \mbox{ -- } 9100$ \\
59667.5 & 2022-03-29 & Keck I & LRIS & $3200 \mbox{ -- } 10400$ \\
59680.4 & 2022-04-11 & University of Hawai`i 2.2-m & SNIFS & $3400 \mbox{ -- } 9100$ \\
59696.5 & 2022-04-27 & University of Hawai`i 2.2-m & SNIFS & $3600 \mbox{ -- } 9100$ \\
59699.3 & 2022-04-30 & Keck I & LRIS & $3200 \mbox{ -- } 10400$ \\
59759.3 & 2022-06-29 & Keck I & LRIS & $3200 \mbox{ -- } 10400$ \\
\hline \multicolumn{5}{c}{Second Flare}\\ \hline
60328.6 & 2024-01-19 & Keck II & KCWI & $3500 \mbox{ -- } 8900$ \\
60402.5 & 2024-04-02 & University of Hawai`i 2.2-m & SNIFS & $3400 \mbox{ -- } 9100$ \\
60403.4 & 2024-04-03 & University of Hawai`i 2.2-m & SNIFS & $3400 \mbox{ -- } 9100$ \\
60405.3 & 2024-04-05 & University of Hawai`i 2.2-m & SNIFS & $3400 \mbox{ -- } 9100$ \\
60407.3 & 2024-04-07 & Keck II & KCWI & $3500 \mbox{ -- } 8900$ \\
\enddata 
\tablecomments{Modified Julian Day, calendar date, telescope, instrument, and observed wavelength range for each of the spectroscopic observations obtained of ASASSN-22ci for the initial classification and during our follow-up campaign.} 
\label{tab:spectra_log} 
\end{deluxetable*}

\section*{Acknowledgements}

We thank Eric Coughlin and Eliot Quataert for their helpful discussions on multiple/repeating TDEs and Jennifer van Saders for useful discussions on stellar evolution. We additionally thank Jonathan Gelbord for productive conversations on the analysis of the jitter-affected UVOT images.

The Shappee group at the University of Hawai'i is supported with funds from NSF (grants AST-1908952, AST-1911074, \& AST-1920392) and NASA (grants HST-GO-17087, 80NSSC24K0521, 80NSSC24K0490, 80NSSC24K0508, 80NSSC23K0058, \& 80NSSC23K1431). CSK and KZS are supported by NSF grants AST-1907570, 2307385, and 2407206. 

We thank Las Cumbres Observatory and its staff for their continued support of ASAS-SN. ASAS-SN is funded  by Gordon and Betty Moore Foundation grants GBMF5490 and GBMF10501 and  the Alfred P. Sloan Foundation grant G-2021-14192.

This research was supported in part by grant NSF PHY-2309135 to the Kavli Institute for Theoretical Physics (KITP).

Parts of this research were supported by the Australian Research Council Discovery Early Career Researcher Award (DECRA) through project number DE230101069.

CA acknowledges support from NASA through grants from the Space Telescope Science Institute  via programs:  JWST-GO-02114, JWST-GO-02122, JWST-GO-04522, JWST-GO-03726, JWST-GO-04575, JWST-GO-05057, JWST-GO-05290, JWST-GO-06023, JWST-GO-06582, JWST-GO-06677
JWST-GO-06716, HST-AR-17555.008-A.

This material is based upon work supported by the National Science Foundation Graduate Research Fellowship Program under Grant Nos. 1842402 and 2236415. Any opinions, findings, conclusions, or recommendations expressed in this material are those of the author(s) and do not necessarily reflect the views of the National Science Foundation.

This publication makes use of data products from the Wide-field Infrared Survey Explorer, which is a joint project of the University of California, Los Angeles, and the Jet Propulsion Laboratory/California Institute of Technology, funded by the National Aeronautics and Space Administration.

This paper includes data collected by the TESS mission. Funding for the TESS mission is provided by the NASA's Science Mission Directorate.

The LBT is an international collaboration among institutions in the United States, Italy and Germany. LBT Corporation partners are as follows: The University of Arizona on behalf of the Arizona Board of Regents; Istituto Nazionale di Astrofisica, Italy; LBT Beteiligungsgesellschaft, Germany, representing the Max-Planck Society, The Leibniz Institute for Astrophysics Potsdam, and Heidelberg University; The Ohio State University, representing OSU, University of Notre Dame, University of Minnesota, and University of Virginia.

Some of the data presented herein were obtained at the W. M. Keck Observatory, which is operated as a scientific partnership among the California Institute of Technology, the University of California and the National Aeronautics and Space Administration. The Observatory was made possible by the generous financial support of the W. M. Keck Foundation.

The NASA Infrared Telescope Facility is operated by the University of Hawaii under contract 80HQTR19D0030 with the National Aeronautics and Space Administration.

This work is based on observations made by ASAS-SN, ATLAS, UH2.2, IRTF, and Keck. The authors wish to recognize and acknowledge the very significant cultural role and reverence that the summits of Haleakal\=a and Maunakea have always had within the indigenous Hawaiian community.  We are most fortunate to have the opportunity to conduct observations from these mountains.

\bibliography{bibliography}
\bibliographystyle{aasjournal}


\label{lastpage}
\end{document}